\documentclass[12pt]{article}
\usepackage{indentfirst}
\usepackage{amsmath}
\usepackage{amssymb}
\usepackage{setspace}
\usepackage{graphicx}
\usepackage{enumerate}
\usepackage{url}
\usepackage{graphicx}
\usepackage{enumerate}
\usepackage[normalem]{ulem}
\usepackage{xfrac}
\usepackage{xcolor}
\usepackage{pdflscape}
\usepackage{longtable}
\usepackage{caption}
\usepackage{booktabs,multirow,array,pdflscape}

\usepackage{soul}
\usepackage{relsize}
\usepackage{multirow}
\usepackage{natbib}

\newlength{\CapLen}
\AtBeginDocument{\settoheight{\CapLen}{A}}

\begin{document}
  \def\spacingset#1{\renewcommand{\baselinestretch}%
  {#1}\small\normalsize} \spacingset{1}

  \title{\bf Inference on the Reliability Function with Type-I Censored Weibull Data}
  \author{ Bowen Liu\\
  Assistant Professor, Division of Computing, Analytics and Mathematics,\\
  UMKC, Kansas City, USA\\[0.5em]
  Samaradasa Weerahandi$^{*}$\\
  President, X-Techniques, New Jersey, USA\\[0.5em]
  Malwane M. A. Ananda\\
  Professor, Department of Mathematical Sciences, UNLV, Las Vegas, USA\\
  }
  \maketitle
  
  \begin{abstract}
This paper develops generalized confidence intervals for the shape parameter,
scale parameter, and reliability function of the two-parameter Weibull
distribution under Type-I censoring. The proposed method uses the
Weibull-to-Gumbel transformation and equivariant least-squares estimators to
construct generalized pivotal quantities without requiring maximum-likelihood
estimators. The resulting generalized confidence intervals are evaluated under
the original Type-I design. Simulation studies show that the procedure,
implemented with the Kaplan--Meier (KM) and Herd--Johnson (HJ) plotting
estimators, generally attains empirical coverage close to the nominal level
across the parameter settings considered. Comparisons with a previously developed
generalized pivotal method, bootstrap procedures, and asymptotic likelihood-based
intervals show that the proposed method generally achieves near-nominal coverage,
while relative interval lengths vary across scenarios. Two reliability examples
illustrate the practical use of the method for Weibull parameter and reliability
inference.
  \end{abstract}

  \noindent%

  \vfill

  \newpage
  \spacingset{1.9}

  \addtolength{\textheight}{-.3in}%

  \bigskip
  \section{Introduction}
\label{sec:introduction}

  Reliability inference based on parametric distributions has been, and still is,
  an important problem in electrical and mechanical engineering applications. When
  expressed in alternative forms, the problem also arises in survival models
  in clinical research. Lifetime distributions such as Weibull \citep{sarhan2023stress,patil2024estimation,fonseca2024estimation, das2025reliability,zou2025stress, rajput2026discriminating},
  loglogistic \citep{felipe2025robust,das2025reliability}, and lognormal \citep{fernandez2023optimum,rajput2026discriminating, das2025reliability}
  are widely used in modeling reliability problems. Recent developments continue
to expand the scope of reliability methodology through degradation modeling,
accelerated life testing, and heterogeneous lifetime models
\citep{zhuang_modeling_2026, miao_generalized_2026, wu_modeling_2026}, emphasizing the need for
accurate statistical inference under practical testing conditions.

Despite their practical importance, confidence intervals for the reliability
\(S(t)\) at a prespecified time \(t\) under Type-I censoring are often based on asymptotic
approximations, bootstrap procedures, or generalized inference. When the data
are censored and the sample size is small, as is common in reliability
applications, accurate interval estimation becomes considerably more
challenging. \citet{Xiang2015} and \citet{Jia2016} developed generalized pivotal
quantity (GPQ) procedures for Type-I censored Weibull data that use both maximum
likelihood estimators (MLEs) and least-squares estimators (LSEs). This study complements that literature by
developing an alternative GPQ based only on least-squares estimators. The
construction first transforms Weibull data to the Gumbel location--scale family
and then maps the resulting GPQs back to the Weibull scale.

Motivated by the important work of \citet{Zhang2007}, we derive generalized pivotal quantities using equivariant least-squares estimators and a parameter-free representation tailored to the Type-I censoring design.
This provides generalized inference for the two-parameter Weibull model and its reliability function.

\subsection{Model formulation}
To outline the problem, let $X$ denote the lifetime of an electrical or mechanical
component. The reliability function is defined as
\begin{equation}
  S(x)=\Pr(X\ge x)=1-F_X(x),
  \label{S}
\end{equation}
where $F_X(x)$ denotes the cumulative distribution function (CDF) of $X$. For many
engineering applications, $X$ is adequately modeled by a two-parameter Weibull
distribution, while practical life tests are frequently terminated at a
pre-specified censoring time, resulting in Type-I censored observations.
Consequently, constructing accurate confidence intervals for reliability-related
quantities under Type-I censoring is a fundamental yet challenging problem in reliability engineering.

For Weibull-distributed lifetimes, \citet{Krishnamoorthy2009} 
developed a
GPQ approach for inference under Type-II
censoring, where the experiment terminates after a fixed number of failures.
\citet{Xiang2015} and \citet{Jia2016} subsequently developed
GPQ procedures for Type-I censored Weibull data that use both MLEs and LSEs.
Building on this body of work, the present article develops a complementary GPQ
that depends only on LSEs. Because GPQs are generally not unique, alternative
constructions may exhibit different finite-sample coverage and interval-length
properties. Our construction uses a monotone transformation of Weibull data to
the Gumbel location--scale family.

The key idea is to exploit the equivariance of the LSEs in the
Gumbel location--scale model. For a realized failure count \(R=r\), the
ordered failure times admit a representation in terms of Uniform\((0,1)\) order
statistics. Under Type-I censoring, \(C>0\) is the common, predetermined time
at which the experiment ends. A unit with lifetime \(X>C\) is right-censored
at \(C\), so the censoring probability is \(p_c=\Pr(X>C)\).
Correspondingly, we define \(p=\Pr(X\le C)=1-p_c\) as the probability of
observing a failure by \(C\).
The failure probability \(p\) is recovered through an estimating
equation, after which the generalized pivotal quantities are obtained by
substitution. The resulting intervals are evaluated under the original Type-I experiment, where \(R\) remains random. The proposed procedure exhibits favorable finite-sample
coverage--length behavior in the scenarios we examine. Its principal contributions are:

\begin{enumerate}
\item We develop LSE-based GPQs for the Weibull distribution under
Type-I censoring using a parameter-free order-statistic representation
associated with the realized number of failures.

\item We establish the GPQ properties for the Weibull shape, scale, and
reliability parameters under the stated regularity conditions.

\item We conduct an extensive numerical study, including comprehensive simulation experiments and real-data reliability applications, to compare the proposed procedure with the generalized pivotal, bootstrap, and likelihood-based methods under a broad range of Type-I censoring scenarios.
\end{enumerate}

The remainder of the article is organized as follows. Section~\ref{Sec:2}
reviews the Weibull distribution theory, plotting estimators, and
log-location--scale representation. Section~\ref{sec:GPQ} develops the
GPQ methodology. Section~\ref{Sec:4} presents an extensive Monte Carlo simulation
study comparing the proposed procedure with the existing generalized pivotal, bootstrap, and maximum-likelihood-based methods under a variety of Type-I
censoring scenarios. Section~\ref{sec:examples}
illustrates the proposed methodology using two numerical examples. Concluding
remarks are provided in Section~\ref{sec: discussion}.

\subsection{The approach}

The purpose of this section is to present the generalized-inference
framework used in this paper for reliability-related quantities under the
two-parameter Weibull model by employing the GPQ methodology of
\citet{Weerahandi1993,Weerahandi1995,Weerahandi2004}. Briefly, let
$\mathbf{X}=(X_1,\ldots,X_n)$ denote a random sample vector from a distribution
having the parameter vector $\boldsymbol{\theta}$, and let
$\mathbf{x}=(x_1,\ldots,x_n)$ denote its observed realization. A random variable
$T(\mathbf{X},\mathbf{x},\boldsymbol{\theta})$ is called a GPQ for a component
$\theta$ of $\boldsymbol{\theta}$ if it satisfies
the following two conditions: 
\begin{enumerate}
    \item 
$T(\mathbf{x},\mathbf{x},\boldsymbol{\theta})=\theta$, and 
\item the distribution
of $T(\mathbf{X},\mathbf{x},\boldsymbol{\theta})$ is free of unknown parameters. 
\end{enumerate}
The observed sample is held fixed when evaluating reference probabilities.
For example, the classical and generalized tail events
\[
\{T_0(\mathbf{X})>T_0(\mathbf{x})\},
\qquad
\{T(\mathbf{X},\mathbf{x},\boldsymbol{\theta})>
T(\mathbf{x},\mathbf{x},\boldsymbol{\theta})\},
\]
where \(T_0\) denotes a classical statistic, are fixed subsets of the
reference sample space for fixed \(\mathbf{x}\) and parameter values.
Their probabilities are numerical values evaluated under the reference
distribution of \(\mathbf{X}\).

For interval estimation, choose a region \(C_{1-\gamma}\) in the range
of the GPQ such that
\[
\Pr\{T(\mathbf{X},\mathbf{x},\boldsymbol{\theta})\in C_{1-\gamma}\}
=1-\gamma.
\]
The observed-value property
\(T(\mathbf{x},\mathbf{x},\boldsymbol{\theta})=\theta\) therefore identifies
\(C_{1-\gamma}\) as a \(100(1-\gamma)\%\) generalized confidence region
for \(\theta\) \citep[Chapter~6]{Weerahandi1995}.
We refer to this generalized confidence level as ``exact'' because the probability
in the statement equals exactly \(1-\gamma\) under the GPQ reference distribution.
In this paper, the required GPQ quantiles are obtained
numerically using Monte Carlo sampling.

In general, the proposed framework consists of three main steps. 
First, identify a suitable
set of statistics that adequately summarize the information in the observed sample. Whenever available, minimum sufficient statistics are preferred; otherwise, widely accepted estimators with desirable properties,
such as the LSEs or MLEs, may be employed. Second, derive generalized
pivotal quantities for the model parameters directly from the underlying
sampling distribution or through an appropriate 
standardized transformation.
For the Weibull distribution considered here, a convenient 
location-scale representation is obtained through the logarithmic transformation.
The present article develops the GPQ construction for the Weibull model.
Finally, generalized confidence intervals for the parameters of interest, including
the reliability function, are obtained from the quantiles of the corresponding GPQ distributions.

The remainder of this section illustrates these ideas for the two-parameter Weibull distribution under Type-I censoring. 
Unlike Type-II censoring, the number of observed failures is random. 
Existing GPQ procedures for Type-I
censored data \citep{Xiang2015,Jia2016} provide relevant benchmarks based on GPQs that depend on
both MLEs and LSEs.
Our approach uses the realized
failure count to obtain a Uniform order-statistic representation of the observed failures and then solves for the probability of observing a failure by \(C\), \(p=\Pr(X\le C)=1-p_c\), within each GPQ draw.
This representation supplies the required parameter-free reference variables
while retaining the original fixed censoring time. The final procedure is evaluated under the Type-I design with a random number of failures.

The required parameter-free representation is established for each admissible
realized failure count. The resulting methodology is subsequently
compared with the WLMA generalized pivotal procedure of \citet{Xiang2015}, LSE/MLE-based
bootstrap intervals, and asymptotic MLE intervals through an extensive
simulation study.

  \section{Distribution Theory for the Weibull Distribution}
\label{Sec:2}

Assume that the lifetime distribution of an electrical or mechanical component
follows a Weibull distribution. Consider the problem of making inferences on the
reliability of such components when the observed data may be subject to Type-I censoring. Existing 
generalized pivotal approaches for Weibull parameters and
reliability include MLE/LSE-based procedures developed under their stated
regularity conditions \citep{Xiang2015,Jia2016}. The MLE construction builds on
distributional results of \citet{Thoman1969}, whereas the LSE construction
requires separate probability-plot arguments. The present article develops an
LSE-based GPQ construction tailored to Type-I censoring.

Let
\[
X_1,X_2,\ldots,X_n
\]
be a random sample from a Weibull distribution \(W(\alpha,\theta)\), where
\(\alpha>0\) is the shape parameter and \(\theta>0\) is the scale parameter.
The probability density function of each \(X_i\) is

\begin{equation}
\label{eqn:weibull_pdf}
f(x;\alpha,\theta)
=
\frac{\alpha}{\theta}
\left(
\frac{x}{\theta}
\right)^{\alpha-1}
\exp\left\{
-
\left(
\frac{x}{\theta}
\right)^\alpha
\right\},
\qquad x>0,
\end{equation}

and the cumulative distribution function is

\begin{equation}
\label{eq:weibull-cdf}
F(x;\alpha,\theta)
= 1- \exp\left\{-
\left( 
\frac{x}{\theta}
\right)^\alpha \right\},
\qquad x>0.
\end{equation}

Suppose the experiment is conducted under Type-I censoring with a common,
predetermined censoring time \(C>0\). The observed data consist of

\[
Y_i=\min(X_i,C),
\qquad
\delta_i=I(X_i\le C),
\qquad
i=1,\ldots,n.
\]

The quantity

\[
R=\sum_{i=1}^{n}\delta_i
\]

denotes the random number of failures observed before the censoring time. Its
distribution depends on \(C\) and the unknown Weibull parameters. For an
observed realization \(R=r\), the ordered failure times are denoted by

\[
X_{1:r}<X_{2:r}<\cdots<X_{r:r}.
\]

This notation is used throughout the remainder of the paper and the Appendix.

\subsection{LSEs of Parameters}

\citet{Zhang2007} discuss estimation of the Weibull shape and
scale parameters by fitting a straight line on Weibull probability paper. Since
\[
\log(x)
\quad\text{and}\quad
\log\{-\log(1-F(x))\}
=
\log\{-\log(S(x))\}
\]
are linearly related under the Weibull model, the regression requires a plotting
estimate of \(F\), or equivalently of \(S=1-F\), at each observed failure time.
The regression may be fitted in either direction; following \citet{Zhang2007},
we regress \(\log(x)\) on \(\log\{-\log(\widehat S(x))\}\), which generally performs
better than the reverse formulation.

In the case of complete data (\(R=n\)), the plotting estimates are usually obtained from
fixed plotting positions that approximate the cumulative distribution function.
Common choices include the Gringorten plotting positions \citep{Gringorten1963}

\[
p_i=\frac{i-0.44}{n+0.12},
\]
and the median-rank plotting positions

\[
p_i=\frac{i-0.30}{n+0.40},
\qquad i=1,\ldots,n,
\]
recommended by \citet{Zhang2007}. We use the latter throughout the complete-data
simulation study.

For Type-I censored data, some lifetimes are known only to exceed \(C\), so
complete-data plotting positions are no longer appropriate. We instead obtain
the plotting estimates from a nonparametric estimator of the reliability function.
The Kaplan--Meier (KM) product-limit estimator is the most widely used choice
\citep{KaplanMeier1958}.
Let \(\tau_1<\cdots<\tau_J\) denote the distinct ordered observed failure times, let
\(d_j\) be the number of failures at \(\tau_j\), and let \(n_j\) be the number
of units at risk just before \(\tau_j\). Then
\[
\widehat S_{\mathrm{KM}}(t)
=
\prod_{\tau_j\le t}
\left(1-\frac{d_j}{n_j}\right),
\qquad
\widehat F_{\mathrm{KM}}(t)=1-\widehat S_{\mathrm{KM}}(t).
\]
This form accommodates tied failures through \(d_j\). If failures and censoring
occur at the same recorded time, units censored at that time remain in the risk
set when the failures at \(\tau_j\) are counted. The Herd--Johnson (HJ) estimator provides an
alternative and was recommended by \citet{Zhang2007}. Both KM
and HJ provide estimated reliability values from which the Weibull
probability-plot coordinates are calculated.

Let \(a\) and \(b\) denote the intercept and slope from the fitted regression,
respectively. The corresponding LSEs of the Weibull parameters are

\[
\widehat{\alpha}
=
\frac{1}{b},
\qquad
\widehat{\theta}
=
\exp(a).
\]

Additional details regarding the computation of the KM and HJ estimators,
plotting positions, and the resulting LSEs can be found in
\citet{Zhang2007} and \citet{Xiang2015}.

\subsection{Log-Location-Scale Representation}
The Weibull distribution admits a convenient log-transformation that converts it
into a location--scale family. To see this, define

\[
Y=\log(X),
\qquad
\nu=\log(\theta),
\qquad
\sigma=\frac{1}{\alpha}.
\]
Then
\[
Y=\nu+\sigma Z,
\]
where \(Z\) follows the standard minimum Gumbel distribution with cumulative
distribution function
\begin{equation}
\label{eq:gumbel-cdf}
F_0(z)
=
1-\exp\{-\exp(z)\},
\qquad
-\infty<z<\infty.
\end{equation}
Consequently,

\begin{equation}
\label{eq:gumbel-location-scale}
F_Y(y;\nu,\sigma)
=
1-
\exp
\left[
-\exp
\left\{
\frac{y-\nu}{\sigma}
\right\}
\right],
\end{equation}
where \(-\infty<y<\infty\).
Thus, the logarithm of a Weibull random variable belongs to a location--scale
family with location parameter \(\nu\) and scale parameter \(\sigma\). The LSE procedures described above are equivalent to fitting a
 simple linear regression model in the transformed Gumbel scale.

This representation is particularly useful since the standardized variable

\[
Z
=
\frac{Y-\nu}{\sigma}
\]
has a distribution free of unknown parameters. The theoretical properties of
this transformation play a central role in the inferential procedures developed
later in the paper and are established formally in Appendix~A.

    \section{Generalized Pivotal Quantities for Weibull Parameters}
\label{sec:GPQ}
Building on the log-location--scale representation in
Section~\ref{Sec:2}, we develop GPQs for the Weibull parameters. We use
\(Y=\log(X)=\nu+\sigma Z\), where \(\nu=\log(\theta)\),
\(\sigma=1/\alpha\), and \(Z\) has the standard minimum Gumbel distribution
defined in Section~\ref{Sec:2}.

The detailed theoretical justification of the GPQ construction is provided in
Appendix~A. Here we summarize the resulting pivotal quantities used for
inference.

\subsection{Complete-Data GPQs}

Suppose first that all failures are completely observed (i.e., 
\(R=n\)).
Let
\[
w_i
=
\log\{-\log(1-p_i)\},
\qquad
i=1,\ldots,n,
\]
where \(p_i\) denotes the chosen plotting position.
Here \(y_{i:n}\) denotes the \(i\)th ordered log failure time in the complete
sample of size \(n\), and \(Z_{i:n}\) denotes the corresponding order statistic
of the reference sample.

The LSEs of the Gumbel scale and location parameters are
\begin{equation}
\label{eq:lse-complete}
\widehat{\sigma}
=
\frac{
\sum_{i=1}^{n}
(w_i-\bar w)
(y_{i:n}-\bar y)
}
{
\sum_{i=1}^{n}
(w_i-\bar w)^2
},
\qquad
\widehat{\nu}
=
\bar y-\widehat{\sigma}\bar w.
\end{equation}

Let
\[
Z_1,\ldots,Z_n
\stackrel{\mathrm{iid}}{\sim}
F_0,
\]
and define
\[
\widehat{\sigma}(\mathbf Z)
=
\frac{
\sum_{i=1}^{n}
(w_i-\bar w)
(Z_{i:n}-\bar Z)
}
{
\sum_{i=1}^{n}
(w_i-\bar w)^2
},
\]
and
\[
\widehat{\nu}(\mathbf Z)
=
\bar Z-\bar w\,\widehat{\sigma}(\mathbf Z).
\]

As shown in Appendix~A, if \(\widehat{\sigma}_{\mathrm{obs}}\) and
\(\widehat{\nu}_{\mathrm{obs}}\) denote the LSEs calculated from the observed
complete sample, then

\begin{equation}
\label{eq:Gsigma-complete-main}
G_{\sigma}
=
\frac{
\widehat{\sigma}_{\mathrm{obs}}
}
{
\widehat{\sigma}(\mathbf Z)
}
\end{equation}
and
\begin{equation}
\label{eq:Gnu-complete-main}
G_{\nu}
=
\widehat{\nu}_{\mathrm{obs}}
-
G_{\sigma}
\widehat{\nu}(\mathbf Z)
\end{equation}
are generalized pivotal quantities for \(\sigma\) and \(\nu\), respectively.

Using the substitution property of GPQs \citep{Weerahandi2004}, the
corresponding GPQs for the Weibull parameters are
\begin{equation}
\label{eq:G-alpha-theta}
G_{\alpha}
=
\frac{1}{G_{\sigma}},
\qquad
G_{\theta}
=
\exp(G_{\nu}).
\end{equation}

\subsection{Type-I Censored Data}
Now suppose the data are obtained under a Type-I censoring scheme with
censoring time \(C\). Let
\[
R=\sum_{i=1}^{n}\delta_i
\]
denote the number of observed failures. Once the Type-I censored sample
is observed, its failure count \(R=r\) is known and is held fixed in the
reference-distribution derivation.
At least two observed failures are required to define the least-squares slope
in the probability-plot regression; when
\(r=n\), the complete-data construction is used.

In this setting, the plotting quantities are obtained from a nonparametric
estimate of the distribution function, such as the KM or HJ estimator. The
resulting LSEs are computed from the ordered observed failure times
\[
X_{1:r}<X_{2:r}<\cdots<X_{r:r}.
\]

Let
\[
c=\log(C).
\]
As shown in Appendix~A, fixing \(R=r\) yields a representation of the ordered
standardized failure times in terms of Uniform\((0,1)\) order statistics.
Specifically,
\[
\mathbf U^*
=
(U_{1:r}^*,\ldots,U_{r:r}^*)
\]
denotes a generated set of Uniform\((0,1)\) order statistics and
\[
q(u)=\log\{-\log(1-u)\}
\]
denotes the quantile function of the standard minimum Gumbel distribution.

The unknown probability of failure by time \(C\), \(p=\Pr(X\le C)\), is recovered through the estimating equation
\begin{equation}
\label{eq:ptilde-main}
K_{\mathrm{obs}}
=
\frac{
q(\widetilde p)
-
\widehat{\nu}(\mathbf U^*,\widetilde p)
}
{
\widehat{\sigma}(\mathbf U^*,\widetilde p)
},
\end{equation}
where
\[
K_{\mathrm{obs}}
=
\frac{
c-\widehat{\nu}_{r,\mathrm{obs}}
}
{
\widehat{\sigma}_{r,\mathrm{obs}}
}.
\]

The solution \(\widetilde p\in(0,1)\) incorporates the observed censoring time
through \(c=\log(C)\). The development assumes the regularity conditions that
\[
\widehat{\sigma}(\mathbf U^*,\widetilde p)\neq 0,
\]
and that Equation (\ref{eq:ptilde-main}) admits a unique solution
\[
\widetilde p\in(0,1).
\]

The resulting GPQs for the Gumbel scale and location parameters are
\begin{equation}
\label{eq:Gsigma-censored-main}
G_{\sigma}
=
\frac{
\widehat{\sigma}_{r,\mathrm{obs}}
}
{
\widehat{\sigma}(\mathbf U^*,\widetilde p)
},
\end{equation}
and
\begin{equation}
\label{eq:Gnu-censored-main}
G_{\nu}
=
c-G_{\sigma}q(\widetilde p).
\end{equation}

The detailed derivation and theoretical justification are provided in
Appendix~A. The corresponding GPQs for the original Weibull parameters are
\[
G_{\alpha}
=
\frac{1}{G_{\sigma}},
\qquad
G_{\theta}
=
\exp(G_{\nu}).
\]

Generalized confidence intervals are obtained by repeated 
generation of the
corresponding GPQs and using the appropriate empirical quantiles of the
resulting Monte Carlo distribution.

\subsection{Inference on the Reliability Function}

For a Weibull random variable \(X\sim W(\alpha,\theta)\), the reliability
at a prespecified time \(t>0\) is

\begin{equation}
\label{eq:survival}
S(t)
=
\Pr(X>t)
=
\exp\left\{
-
\left(
\frac{t}{\theta}
\right)^\alpha
\right\}.
\end{equation}

Using the log-location-scale representation introduced in
Section~\ref{Sec:2}, the reliability function may equivalently be expressed as
\begin{equation}
\label{eq:survival-gumbel}
S(t)
=
\exp\left[
-
\exp\left\{
\frac{\log(t)-\nu}{\sigma}
\right\}
\right],
\end{equation}
where
\[
\nu=\log(\theta),
\qquad
\sigma=\frac{1}{\alpha}.
\]

Therefore, once generalized pivotal quantities for \(\nu\) and \(\sigma\) (or
equivalently for \(\theta\) and \(\alpha\)) have been obtained, a GPQ for the
reliability at time \(t\) follows directly from the substitution property of
generalized pivotal quantities \citep{Weerahandi2004}. Specifically,
\begin{equation}
\label{eq:GPQ-reliability}
G_{S(t)}
=
\exp\left[
-
\left(
\frac{t}{G_{\theta}}
\right)^{G_{\alpha}}
\right].
\end{equation}

Equivalently, in terms of the Gumbel location-scale parameters,
\begin{equation}
\label{eq:GPQ-reliability-gumbel}
G_{S(t)}
=
\exp\left[
-
\exp\left\{
\frac{\log(t)-G_{\nu}}
{G_{\sigma}}
\right\}
\right].
\end{equation}

A generalized confidence interval for \(S(t)\) is obtained from the empirical
quantiles of the Monte Carlo distribution of \(G_{S(t)}\). For a confidence
level of \(100(1-\gamma)\%\), the lower and upper confidence limits are given
by the \(\gamma/2\) and \(1-\gamma/2\) empirical quantiles of the generated values of
\(G_{S(t)}\), respectively. Generalized hypothesis testing procedures concerning the
reliability at a prespecified time may also be performed using the
corresponding generalized confidence intervals.

The procedure for obtaining the generalized confidence intervals is summarized
in the following algorithm.

\begingroup
\singlespacing
\small
\noindent\textbf{Algorithm 1 (GPQ-based confidence intervals).}
Given \(n\) lifetimes with \(2\le r\le n\) observed failures, choose \(B\) Monte Carlo
draws, a prespecified time \(t>0\), and confidence level \(100(1-\gamma)\%\). For \(r<n\), specify
\(C\) and the KM or HJ estimator. Assume the regularity conditions in
Section 3.2 and Appendix A.

\begin{enumerate}
\setlength{\itemsep}{3pt}
\setlength{\parsep}{0pt}
\item \textbf{Compute the observed LSEs.}
Order the observed failure times and take their logarithms. For complete
data (\(r=n\)), use the median-rank positions
\(p_i=(i-0.30)/(n+0.40)\); for \(r<n\), use the selected KM or HJ plotting
estimates. Set \(w_i=q(p_i)\), with \(q(u)=\log\{-\log(1-u)\}\), and
regress the ordered log failure times on \(w_i\) to obtain the slope and
intercept LSEs. Keep these scores fixed in all reference regressions.
For \(r<n\), also compute \(c=\log C\) and
\(K_{\mathrm{obs}}=(c-\widehat\nu_{r,\mathrm{obs}})/\widehat\sigma_{r,\mathrm{obs}}\).

\item \textbf{Generate a reference sample for draw \(b\).}
For complete data, generate \(n\) independent standard minimum Gumbel
variables, order them, and regress them on the same \(w_i\) to obtain
\(\widehat\sigma(\mathbf Z^{(b)})\) and
\(\widehat\nu(\mathbf Z^{(b)})\).
No probability-matching equation is solved for complete data.

For \(r<n\), generate and order \(r\) independent Uniform\((0,1)\)
variables to form \(\mathbf U^{*(b)}\). For each candidate \(p\in(0,1)\),
regress \(q(pU_{j:r}^{*(b)})\), \(j=1,\ldots,r\), on the fixed \(w_j\)
to obtain \(\widehat\sigma(\mathbf U^{*(b)},p)\) and
\(\widehat\nu(\mathbf U^{*(b)},p)\). Solve
Equation~\eqref{eq:ptilde-main} for \(\widetilde p^{(b)}\) using a
sufficiently fine grid search followed by numerical root refinement.
Keep the entire vector \(\mathbf U^{*(b)}\) unchanged throughout this
search and the subsequent GPQ calculation.

\item \textbf{Compute the Gumbel-parameter GPQs.}
For complete data, use
Equations~\eqref{eq:Gsigma-complete-main}--\eqref{eq:Gnu-complete-main}.
For \(r<n\), substitute the same \(\mathbf U^{*(b)}\) and its solution
\(\widetilde p^{(b)}\) into
Equations~\eqref{eq:Gsigma-censored-main}--\eqref{eq:Gnu-censored-main}.
This gives \(G_\sigma^{(b)}\) and \(G_\nu^{(b)}\).

\item \textbf{Transform each draw to the quantities of interest.}
Compute and store
\[
G_\alpha^{(b)}=\frac{1}{G_\sigma^{(b)}},\qquad
G_\theta^{(b)}=\exp(G_\nu^{(b)}),\qquad
G_{S(t)}^{(b)}=
\exp\left[-\left\{\frac{t}{G_\theta^{(b)}}\right\}^{G_\alpha^{(b)}}\right].
\]

\item \textbf{Obtain empirical percentiles.}
Repeat Steps 2--4 independently for \(b=1,\ldots,B\).
For each \(\eta\in\{\alpha,\theta,S(t)\}\), let
\(\widehat Q_\eta(a)\) denote the empirical \(a\)-quantile of
\(G_\eta^{(1)},\ldots,G_\eta^{(B)}\). The \(100(1-\gamma)\%\) interval is
\([\widehat Q_\eta(\gamma/2),\widehat Q_\eta(1-\gamma/2)]\).
For 95\% intervals, use the 2.5th and 97.5th percentiles.
Compute reliability percentiles from the reliability draws themselves,
not from separate parameter-interval endpoints.
\end{enumerate}
\endgroup

  \section{Simulations}
\label{Sec:4}

We assess finite-sample performance first for complete data and then under Type-I censoring. For the nominal 95\% confidence intervals, empirical coverage
probability measures calibration, while average interval length measures
precision.

\subsection{Complete-data scenarios}
\label{subsec:complete-data}

Complete samples were generated from \(W(\alpha,\theta)\) for
\(n\in\{10,20,50\}\), \(\alpha\in\{2,5\}\), and
\(\theta\in\{1,2,5\}\). We assess the coverage probabilities and average
lengths of confidence intervals for \(S(t)\) at the prespecified times
\(t=1\) and \(t=2\).
The proposed generalized location--scale approach (GLA), implemented with the
median-rank plotting positions, was compared with the WLMA generalized pivotal
procedure of \citet{Xiang2015}, Bootstrap--LSE,
Bootstrap--MLE, and asymptotic MLE. These comparators represent
generalized inference, LSE/MLE-based resampling, and conventional
likelihood-based inference. Bootstrap intervals used percentile limits, and
asymptotic reliability intervals used the complementary log--log
transformation. Each scenario used \(2{,}000\) Monte Carlo replications and
\(1{,}000\) GPQ or bootstrap draws per replication.
The shape and scale comparisons in Tables~\ref{tab:complete_scale_results}
and~\ref{tab:complete_shape_results} used independently generated samples
across parameter settings.

Tables~\ref{tab:complete_scale_results}--\ref{tab:complete_reliability_results}
report the results for \(\alpha\), \(\theta\), and \(S(t)\). Across the three
sample sizes, interval lengths decreased as \(n\) increased and the differences
among methods became less pronounced. GLA generally attained near-nominal
coverage with competitive interval lengths. The remaining procedures displayed
different balances between coverage and interval length, particularly for the
smallest samples. Thus, the complete-data results show the expected
coverage--length trade-off rather than a uniformly best procedure.

\begin{table}[htbp]
  \centering
  \caption{Average interval length and coverage probability for the scale parameter $\theta$ for 95\% confidence intervals (2,000 Monte Carlo replications; 1,000 GPQ or bootstrap draws where applicable)}
  \label{tab:complete_scale_results}
  \small
  \resizebox{\textwidth}{!}{%
  \begin{tabular}{|c|c|c|c|c|c|c|c|c|c|c|c|c|}
    \hline
    $n$ & $\alpha$ & $\theta$ & \multicolumn{2}{c|}{WLMA} & \multicolumn{2}{c|}{GLA} & \multicolumn{2}{c|}{Bootstrap--LSE} & \multicolumn{2}{c|}{Bootstrap--MLE} & \multicolumn{2}{c|}{Asymptotic MLE} \\
    \cline{4-13}
         &          &          & Length & Coverage & Length & Coverage & Length & Coverage & Length & Coverage & Length & Coverage \\
    \hline
    \multirow{6}{*}{10} & 2 & 1   & 1.989 & 1.000 & 0.814 & 0.950 & 0.683 & 0.917 & 0.584 & 0.887 & 0.603 & 0.896 \\
     & 2 & 2   & 3.968 & 1.000 & 1.646 & 0.944 & 1.381 & 0.915 & 1.178 & 0.888 & 1.215 & 0.892 \\
     & 2 & 5   & 9.990 & 1.000 & 4.064 & 0.949 & 3.422 & 0.921 & 2.929 & 0.892 & 3.025 & 0.896 \\
     & 5 & 1   & 1.986 & 1.000 & 0.311 & 0.951 & 0.272 & 0.920 & 0.236 & 0.898 & 0.238 & 0.900 \\
     & 5 & 2   & 3.964 & 1.000 & 0.634 & 0.950 & 0.554 & 0.924 & 0.479 & 0.892 & 0.482 & 0.904 \\
     & 5 & 5   & 9.934 & 1.000 & 1.562 & 0.952 & 1.367 & 0.929 & 1.184 & 0.897 & 1.191 & 0.895 \\
    \hline
    \multirow{6}{*}{20} & 2 & 1   & 1.143 & 1.000 & 0.516 & 0.956 & 0.482 & 0.937 & 0.437 & 0.929 & 0.445 & 0.936 \\
     & 2 & 2   & 2.292 & 1.000 & 1.035 & 0.950 & 0.966 & 0.938 & 0.877 & 0.931 & 0.894 & 0.934 \\
     & 2 & 5   & 5.757 & 1.000 & 2.551 & 0.952 & 2.381 & 0.945 & 2.171 & 0.934 & 2.210 & 0.934 \\
     & 5 & 1   & 1.148 & 1.000 & 0.204 & 0.949 & 0.193 & 0.938 & 0.176 & 0.929 & 0.177 & 0.928 \\
     & 5 & 2   & 2.292 & 1.000 & 0.405 & 0.956 & 0.383 & 0.945 & 0.350 & 0.933 & 0.352 & 0.931 \\
     & 5 & 5   & 5.742 & 1.000 & 1.013 & 0.955 & 0.959 & 0.945 & 0.877 & 0.933 & 0.881 & 0.934 \\
    \hline
    \multirow{6}{*}{50} & 2 & 1   & 0.647 & 1.000 & 0.308 & 0.951 & 0.302 & 0.945 & 0.285 & 0.941 & 0.288 & 0.944 \\
     & 2 & 2   & 1.289 & 1.000 & 0.613 & 0.934 & 0.602 & 0.929 & 0.568 & 0.925 & 0.574 & 0.929 \\
     & 2 & 5   & 3.224 & 1.000 & 1.532 & 0.950 & 1.505 & 0.946 & 1.418 & 0.942 & 1.434 & 0.946 \\
     & 5 & 1   & 0.644 & 1.000 & 0.122 & 0.954 & 0.120 & 0.952 & 0.114 & 0.944 & 0.114 & 0.944 \\
     & 5 & 2   & 1.289 & 1.000 & 0.244 & 0.943 & 0.241 & 0.944 & 0.228 & 0.941 & 0.229 & 0.939 \\
     & 5 & 5   & 3.224 & 1.000 & 0.608 & 0.950 & 0.600 & 0.951 & 0.570 & 0.938 & 0.573 & 0.943 \\
    \hline
  \end{tabular}%
  }
\end{table}

\begin{table}[htbp]
  \centering
  \caption{Average interval length and coverage probability for the shape parameter $\alpha$ for 95\% confidence intervals (2,000 Monte Carlo replications; 1,000 GPQ or bootstrap draws where applicable)}
  \label{tab:complete_shape_results} \small
  \resizebox{\textwidth}{!}{%
  \begin{tabular}{|c|c|c|c|c|c|c|c|c|c|c|c|c|}
    \hline
    $n$ & $\alpha$ & $\theta$ & \multicolumn{2}{c|}{WLMA} & \multicolumn{2}{c|}{GLA} & \multicolumn{2}{c|}{Bootstrap--LSE} & \multicolumn{2}{c|}{Bootstrap--MLE} & \multicolumn{2}{c|}{Asymptotic MLE} \\
    \cline{4-13}
    & & & Length & Coverage & Length & Coverage & Length & Coverage & Length & Coverage & Length & Coverage \\
    \hline
    \multirow{6}{*}{10} & 2 & 1   & 2.668 & 0.946 & 2.668 & 0.946 & 2.802 & 0.950 & 3.130 & 0.818 & 2.419 & 0.897 \\
     & 2 & 2   & 2.662 & 0.943 & 2.662 & 0.943 & 2.797 & 0.942 & 3.094 & 0.826 & 2.389 & 0.906 \\
     & 2 & 5   & 2.677 & 0.951 & 2.677 & 0.951 & 2.813 & 0.953 & 3.126 & 0.816 & 2.401 & 0.909 \\
     & 5 & 1   & 6.763 & 0.946 & 6.763 & 0.946 & 7.083 & 0.944 & 7.853 & 0.825 & 6.053 & 0.902 \\
     & 5 & 2   & 6.642 & 0.949 & 6.642 & 0.949 & 6.979 & 0.948 & 7.745 & 0.838 & 5.973 & 0.909 \\
     & 5 & 5   & 6.723 & 0.954 & 6.723 & 0.954 & 7.069 & 0.957 & 7.863 & 0.820 & 6.063 & 0.898 \\
    \hline
    \multirow{6}{*}{20} & 2 & 1   & 1.800 & 0.945 & 1.800 & 0.945 & 1.787 & 0.947 & 1.704 & 0.890 & 1.510 & 0.925 \\
     & 2 & 2   & 1.797 & 0.946 & 1.797 & 0.946 & 1.786 & 0.946 & 1.701 & 0.884 & 1.508 & 0.918 \\
     & 2 & 5   & 1.826 & 0.953 & 1.826 & 0.953 & 1.814 & 0.948 & 1.723 & 0.884 & 1.527 & 0.923 \\
     & 5 & 1   & 4.522 & 0.943 & 4.522 & 0.943 & 4.501 & 0.944 & 4.285 & 0.876 & 3.796 & 0.921 \\
     & 5 & 2   & 4.536 & 0.950 & 4.536 & 0.950 & 4.501 & 0.951 & 4.308 & 0.885 & 3.820 & 0.925 \\
     & 5 & 5   & 4.560 & 0.944 & 4.560 & 0.944 & 4.521 & 0.944 & 4.303 & 0.876 & 3.814 & 0.923 \\
    \hline
    \multirow{6}{*}{50} & 2 & 1   & 1.128 & 0.951 & 1.128 & 0.951 & 1.105 & 0.950 & 0.941 & 0.932 & 0.899 & 0.947 \\
     & 2 & 2   & 1.129 & 0.952 & 1.129 & 0.952 & 1.107 & 0.947 & 0.942 & 0.924 & 0.901 & 0.945 \\
     & 2 & 5   & 1.132 & 0.954 & 1.132 & 0.954 & 1.108 & 0.947 & 0.942 & 0.906 & 0.902 & 0.939 \\
     & 5 & 1   & 2.835 & 0.946 & 2.835 & 0.946 & 2.775 & 0.943 & 2.366 & 0.923 & 2.262 & 0.944 \\
     & 5 & 2   & 2.831 & 0.954 & 2.831 & 0.954 & 2.772 & 0.949 & 2.364 & 0.907 & 2.263 & 0.929 \\
     & 5 & 5   & 2.842 & 0.946 & 2.842 & 0.946 & 2.782 & 0.944 & 2.358 & 0.917 & 2.253 & 0.944 \\
    \hline
  \end{tabular}%
  }

\end{table}

\begin{table}[htbp]
  \centering
  \caption{Average interval length and coverage probability for the reliability function $S(t)$ for 95\% confidence intervals (2,000 Monte Carlo replications; 1,000 GPQ or bootstrap draws where applicable)}
  \label{tab:complete_reliability_results}
  \small
  \resizebox{\textwidth}{!}{%
  \begin{tabular}{|c|c|c|c|c|c|c|c|c|c|c|c|c|c|}
    \hline
    $n$ & $t$ & $\alpha$ & $\theta$ & \multicolumn{2}{c|}{WLMA} & \multicolumn{2}{c|}{GLA} & \multicolumn{2}{c|}{Bootstrap--LSE} & \multicolumn{2}{c|}{Bootstrap--MLE} & \multicolumn{2}{c|}{Asymptotic MLE} \\
    \cline{5-14}
         &     &          &          & Length & Coverage & Length & Coverage & Length & Coverage & Length & Coverage & Length & Coverage \\
    \hline
    \multirow{12}{*}{10} & 1 & 2 & 1 & 0.878 & 1.000 & 0.448 & 0.954 & 0.473 & 0.930 & 0.495 & 0.909 & 0.433 & 0.918 \\
     & 1 & 2 & 2 & 0.561 & 0.996 & 0.402 & 0.947 & 0.399 & 0.943 & 0.367 & 0.888 & 0.410 & 0.929 \\
     & 1 & 2 & 5 & 0.201 & 0.979 & 0.207 & 0.956 & 0.192 & 0.957 & 0.125 & 0.857 & 0.204 & 0.933 \\
     & 1 & 5 & 1 & 0.995 & 1.000 & 0.448 & 0.954 & 0.473 & 0.930 & 0.495 & 0.909 & 0.433 & 0.918 \\
     & 1 & 5 & 2 & 0.489 & 0.999 & 0.187 & 0.955 & 0.172 & 0.957 & 0.107 & 0.855 & 0.183 & 0.933 \\
     & 1 & 5 & 5 & 0.033 & 0.989 & 0.029 & 0.954 & 0.024 & 0.950 & 0.007 & 0.841 & 0.023 & 0.936 \\
     & 2 & 2 & 1 & 0.577 & 0.999 & 0.208 & 0.954 & 0.156 & 0.950 & 0.085 & 0.807 & 0.139 & 0.933 \\
     & 2 & 2 & 2 & 0.878 & 1.000 & 0.448 & 0.954 & 0.473 & 0.930 & 0.495 & 0.909 & 0.433 & 0.918 \\
     & 2 & 2 & 5 & 0.423 & 0.992 & 0.352 & 0.949 & 0.341 & 0.949 & 0.289 & 0.878 & 0.357 & 0.927 \\
     & 2 & 5 & 1 & 0.788 & 0.999 & 0.018 & 0.953 & 0.010 & 0.950 & 0.000 & 0.803 & 0.005 & 0.927 \\
     & 2 & 5 & 2 & 0.995 & 1.000 & 0.448 & 0.954 & 0.473 & 0.930 & 0.495 & 0.909 & 0.433 & 0.918 \\
     & 2 & 5 & 5 & 0.215 & 0.999 & 0.117 & 0.954 & 0.106 & 0.954 & 0.054 & 0.854 & 0.111 & 0.935 \\
    \hline
    \multirow{12}{*}{20} & 1 & 2 & 1 & 0.698 & 1.000 & 0.330 & 0.956 & 0.339 & 0.941 & 0.348 & 0.926 & 0.322 & 0.943 \\
     & 1 & 2 & 2 & 0.417 & 0.997 & 0.301 & 0.957 & 0.299 & 0.956 & 0.279 & 0.934 & 0.295 & 0.945 \\
     & 1 & 2 & 5 & 0.143 & 0.981 & 0.138 & 0.957 & 0.136 & 0.954 & 0.096 & 0.929 & 0.126 & 0.949 \\
     & 1 & 5 & 1 & 0.963 & 1.000 & 0.330 & 0.956 & 0.339 & 0.941 & 0.348 & 0.926 & 0.322 & 0.943 \\
     & 1 & 5 & 2 & 0.260 & 1.000 & 0.121 & 0.957 & 0.120 & 0.954 & 0.081 & 0.928 & 0.110 & 0.949 \\
     & 1 & 5 & 5 & 0.013 & 0.989 & 0.010 & 0.953 & 0.010 & 0.953 & 0.004 & 0.918 & 0.007 & 0.949 \\
     & 2 & 2 & 1 & 0.277 & 1.000 & 0.131 & 0.962 & 0.114 & 0.966 & 0.066 & 0.876 & 0.091 & 0.947 \\
     & 2 & 2 & 2 & 0.698 & 1.000 & 0.330 & 0.956 & 0.339 & 0.941 & 0.348 & 0.926 & 0.322 & 0.943 \\
     & 2 & 2 & 5 & 0.320 & 0.993 & 0.260 & 0.952 & 0.256 & 0.955 & 0.224 & 0.931 & 0.251 & 0.944 \\
     & 2 & 5 & 1 & 0.160 & 1.000 & 0.002 & 0.957 & 0.002 & 0.960 & 0.000 & 0.885 & 0.000 & 0.952 \\
     & 2 & 5 & 2 & 0.963 & 1.000 & 0.330 & 0.956 & 0.339 & 0.941 & 0.348 & 0.926 & 0.322 & 0.943 \\
     & 2 & 5 & 5 & 0.116 & 0.997 & 0.067 & 0.952 & 0.067 & 0.953 & 0.038 & 0.925 & 0.057 & 0.951 \\
    \hline
    \multirow{12}{*}{50} & 1 & 2 & 1 & 0.456 & 1.000 & 0.214 & 0.947 & 0.215 & 0.944 & 0.217 & 0.938 & 0.210 & 0.944 \\
     & 1 & 2 & 2 & 0.278 & 1.000 & 0.196 & 0.960 & 0.195 & 0.958 & 0.182 & 0.940 & 0.187 & 0.960 \\
     & 1 & 2 & 5 & 0.089 & 0.983 & 0.081 & 0.956 & 0.082 & 0.951 & 0.063 & 0.937 & 0.070 & 0.958 \\
     & 1 & 5 & 1 & 0.835 & 1.000 & 0.214 & 0.947 & 0.215 & 0.944 & 0.217 & 0.939 & 0.210 & 0.944 \\
     & 1 & 5 & 2 & 0.135 & 1.000 & 0.069 & 0.956 & 0.071 & 0.952 & 0.053 & 0.937 & 0.060 & 0.956 \\
     & 1 & 5 & 5 & 0.004 & 0.986 & 0.003 & 0.955 & 0.004 & 0.947 & 0.002 & 0.934 & 0.002 & 0.952 \\
     & 2 & 2 & 1 & 0.124 & 0.994 & 0.075 & 0.947 & 0.073 & 0.952 & 0.046 & 0.922 & 0.054 & 0.949 \\
     & 2 & 2 & 2 & 0.456 & 1.000 & 0.214 & 0.947 & 0.215 & 0.944 & 0.217 & 0.938 & 0.210 & 0.944 \\
     & 2 & 2 & 5 & 0.215 & 0.995 & 0.167 & 0.963 & 0.167 & 0.958 & 0.148 & 0.939 & 0.155 & 0.962 \\
     & 2 & 5 & 1 & 0.004 & 1.000 & 0.000 & 0.951 & 0.000 & 0.946 & 0.000 & 0.931 & 0.000 & 0.963 \\
     & 2 & 5 & 2 & 0.835 & 1.000 & 0.214 & 0.947 & 0.215 & 0.944 & 0.217 & 0.938 & 0.210 & 0.944 \\
     & 2 & 5 & 5 & 0.056 & 1.000 & 0.034 & 0.955 & 0.035 & 0.947 & 0.023 & 0.936 & 0.028 & 0.954 \\
    \hline
  \end{tabular}%
  }
\end{table}

\subsection{Type-I censored scenarios}
\label{subsec:type1-simulations}
Samples were generated from \(W(\alpha,\theta)\) for
\(n\in\{20,30,50\}\), \(\alpha\in\{2,5\}\), and
\(\theta\in\{1,2,5\}\). For target censoring probability
\(p_c=\Pr(X>C)\in\{0.20,0.30,0.50\}\), the censoring time
\[
C=\theta\{-\log(p_c)\}^{1/\alpha}
\]
was fixed across replications; the realized censoring proportion and number of
failures were therefore random.

The proposed GLA was implemented with both KM and HJ plotting estimates. For
the observed failure count \(R=r\), the auxiliary probability
\(\widetilde p\) was obtained from Equation~\eqref{eq:ptilde-main} using a
sufficiently fine grid search. Each interval used \(1{,}000\) GPQ draws. The
comparison methods were
WLMA \citep{Xiang2015}, Bootstrap--LSE, Bootstrap--MLE, and asymptotic MLE;
bootstrap samples were censored at the same fixed \(C\). We assess the coverage
probabilities and average lengths of confidence intervals for \(S(t)\) at the
prespecified times \(t=1.2\) and \(t=1.5\). Across the parameter settings,
\(S(t)\) ranges from values near zero to values near one, allowing assessment
of interval performance for both interior and near-boundary reliability values. Each
scenario used \(2{,}000\) Monte Carlo replications.

\begin{table}[htbp]
  \centering
  \caption{Average interval length and coverage probability for the shape parameter $\alpha$ using the proposed GLA--KM and GLA--HJ intervals under Type-I censoring (95\% confidence intervals; 2,000 Monte Carlo replications; 1,000 GPQ draws)}
  \label{tab:type1_alpha_balanced}
  \small
  \resizebox{\textwidth}{!}{%
  \begin{tabular}{|c|c|c|c|c|c|c|c|c|}
    \hline
    \multirow{2}{*}{$n$} & \multirow{2}{*}{Parameter setting} & \multirow{2}{*}{Method} & \multicolumn{2}{c|}{20\% censoring} & \multicolumn{2}{c|}{30\% censoring} & \multicolumn{2}{c|}{50\% censoring} \\
    \cline{4-9}
    & & & Length & Coverage & Length & Coverage & Length & Coverage \\
    \hline
    \multirow{12}{*}{20} & \multirow{2}{*}{$\alpha=2,\ \theta=1$} & KM & 2.581 & 0.946 & 2.822 & 0.952 & 3.823 & 0.900 \\
    \cline{3-9}
     &  & HJ & 2.536 & 0.965 & 2.892 & 0.933 & 3.728 & 0.897 \\
    \cline{2-9}
     & \multirow{2}{*}{$\alpha=2,\ \theta=2$} & KM & 2.513 & 0.961 & 2.846 & 0.942 & 3.665 & 0.907 \\
    \cline{3-9}
     &  & HJ & 2.520 & 0.961 & 2.825 & 0.950 & 3.661 & 0.914 \\
    \cline{2-9}
     & \multirow{2}{*}{$\alpha=2,\ \theta=5$} & KM & 2.573 & 0.955 & 2.855 & 0.938 & 3.745 & 0.899 \\
    \cline{3-9}
     &  & HJ & 2.551 & 0.964 & 2.848 & 0.940 & 3.748 & 0.899 \\
    \cline{2-9}
     & \multirow{2}{*}{$\alpha=5,\ \theta=1$} & KM & 6.268 & 0.966 & 7.108 & 0.931 & 9.271 & 0.894 \\
    \cline{3-9}
     &  & HJ & 6.330 & 0.956 & 7.064 & 0.940 & 9.500 & 0.881 \\
    \cline{2-9}
     & \multirow{2}{*}{$\alpha=5,\ \theta=2$} & KM & 6.323 & 0.963 & 7.083 & 0.931 & 9.359 & 0.897 \\
    \cline{3-9}
     &  & HJ & 6.393 & 0.947 & 7.047 & 0.953 & 9.243 & 0.903 \\
    \cline{2-9}
     & \multirow{2}{*}{$\alpha=5,\ \theta=5$} & KM & 6.302 & 0.962 & 7.035 & 0.932 & 9.379 & 0.910 \\
    \cline{3-9}
     &  & HJ & 6.408 & 0.944 & 6.923 & 0.964 & 9.623 & 0.888 \\
    \hline
    \multirow{12}{*}{30} & \multirow{2}{*}{$\alpha=2,\ \theta=1$} & KM & 2.116 & 0.964 & 2.304 & 0.951 & 3.030 & 0.909 \\
    \cline{3-9}
     &  & HJ & 2.100 & 0.959 & 2.339 & 0.961 & 2.992 & 0.911 \\
    \cline{2-9}
     & \multirow{2}{*}{$\alpha=2,\ \theta=2$} & KM & 2.075 & 0.962 & 2.308 & 0.945 & 2.960 & 0.928 \\
    \cline{3-9}
     &  & HJ & 2.100 & 0.960 & 2.323 & 0.961 & 2.977 & 0.911 \\
    \cline{2-9}
     & \multirow{2}{*}{$\alpha=2,\ \theta=5$} & KM & 2.084 & 0.976 & 2.321 & 0.947 & 2.955 & 0.916 \\
    \cline{3-9}
     &  & HJ & 2.081 & 0.957 & 2.331 & 0.958 & 2.990 & 0.915 \\
    \cline{2-9}
     & \multirow{2}{*}{$\alpha=5,\ \theta=1$} & KM & 5.162 & 0.957 & 5.701 & 0.950 & 7.388 & 0.907 \\
    \cline{3-9}
     &  & HJ & 5.243 & 0.963 & 5.719 & 0.959 & 7.517 & 0.891 \\
    \cline{2-9}
     & \multirow{2}{*}{$\alpha=5,\ \theta=2$} & KM & 5.185 & 0.959 & 5.765 & 0.945 & 7.508 & 0.905 \\
    \cline{3-9}
     &  & HJ & 5.171 & 0.954 & 5.689 & 0.960 & 7.383 & 0.913 \\
    \cline{2-9}
     & \multirow{2}{*}{$\alpha=5,\ \theta=5$} & KM & 5.180 & 0.960 & 5.698 & 0.962 & 7.369 & 0.921 \\
    \cline{3-9}
     &  & HJ & 5.217 & 0.953 & 5.808 & 0.942 & 7.428 & 0.913 \\
    \hline
    \multirow{12}{*}{50} & \multirow{2}{*}{$\alpha=2,\ \theta=1$} & KM & 1.647 & 0.957 & 1.819 & 0.947 & 2.286 & 0.921 \\
    \cline{3-9}
     &  & HJ & 1.648 & 0.976 & 1.813 & 0.955 & 2.268 & 0.936 \\
    \cline{2-9}
     & \multirow{2}{*}{$\alpha=2,\ \theta=2$} & KM & 1.646 & 0.964 & 1.844 & 0.948 & 2.244 & 0.928 \\
    \cline{3-9}
     &  & HJ & 1.652 & 0.966 & 1.849 & 0.960 & 2.228 & 0.929 \\
    \cline{2-9}
     & \multirow{2}{*}{$\alpha=2,\ \theta=5$} & KM & 1.649 & 0.967 & 1.817 & 0.955 & 2.274 & 0.931 \\
    \cline{3-9}
     &  & HJ & 1.650 & 0.959 & 1.802 & 0.956 & 2.293 & 0.918 \\
    \cline{2-9}
     & \multirow{2}{*}{$\alpha=5,\ \theta=1$} & KM & 4.096 & 0.961 & 4.548 & 0.958 & 5.717 & 0.930 \\
    \cline{3-9}
     &  & HJ & 4.127 & 0.968 & 4.497 & 0.968 & 5.597 & 0.928 \\
    \cline{2-9}
     & \multirow{2}{*}{$\alpha=5,\ \theta=2$} & KM & 4.131 & 0.959 & 4.538 & 0.963 & 5.700 & 0.923 \\
    \cline{3-9}
     &  & HJ & 4.142 & 0.961 & 4.574 & 0.959 & 5.701 & 0.927 \\
    \cline{2-9}
     & \multirow{2}{*}{$\alpha=5,\ \theta=5$} & KM & 4.134 & 0.957 & 4.552 & 0.958 & 5.612 & 0.940 \\
    \cline{3-9}
     &  & HJ & 4.056 & 0.971 & 4.535 & 0.957 & 5.626 & 0.943 \\
    \hline
  \end{tabular}%
  }

  \begin{minipage}{0.98\textwidth}
  \footnotesize
  \emph{Note.} The fixed censoring time \(C\) was determined by the target censoring proportion, as described in the simulation design. KM denotes the Kaplan--Meier plotting estimator and HJ denotes the Herd--Johnson plotting estimator.
  \end{minipage}
\end{table}

\bigskip

\begin{landscape}
\begin{table}[p]
\centering
\caption{Coverage probability and average interval length for the shape parameter $\alpha$ using the comparison methods under Type-I censoring (95\% confidence intervals; 2,000 Monte Carlo replications; 1,000 GPQ or bootstrap draws where applicable)}
\label{tab:type1_alpha_transposed}
\fontsize{7.2}{8.1}\selectfont
\setlength{\tabcolsep}{1.8pt}
\renewcommand{\arraystretch}{0.94}
\textit{Entries are coverage probability (average interval length).}\par\vspace{3pt}
\begin{tabular}{@{}c*{12}{c}@{}}
\toprule
\multicolumn{13}{c}{\textbf{Panel: $n=20$}} \\
\addlinespace[1pt]
$(\alpha,\theta)$ & \multicolumn{4}{c}{20\% censoring} & \multicolumn{4}{c}{30\% censoring} & \multicolumn{4}{c}{50\% censoring} \\
\cmidrule(lr){2-5}\cmidrule(lr){6-9}\cmidrule(lr){10-13}
 & WLMA & B--LSE & B--MLE & A--MLE & WLMA & B--LSE & B--MLE & A--MLE & WLMA & B--LSE & B--MLE & A--MLE \\
\midrule
$(2,1)$ & 0.905 (2.365) & 0.937 (2.054) & 0.909 (1.990) & 0.938 (1.790) & 0.880 (2.698) & 0.934 (2.322) & 0.914 (2.246) & 0.952 (1.994) & 0.871 (3.629) & 0.951 (3.142) & 0.925 (3.213) & 0.943 (2.555) \\
$(2,2)$ & 0.890 (2.349) & 0.925 (2.043) & 0.916 (1.987) & 0.946 (1.782) & 0.886 (2.689) & 0.938 (2.311) & 0.916 (2.277) & 0.950 (2.002) & 0.875 (3.581) & 0.959 (3.127) & 0.928 (3.196) & 0.962 (2.562) \\
$(2,5)$ & 0.905 (2.324) & 0.931 (2.030) & 0.923 (1.988) & 0.949 (1.785) & 0.895 (2.667) & 0.942 (2.312) & 0.915 (2.265) & 0.939 (1.997) & 0.879 (3.581) & 0.947 (3.109) & 0.931 (3.221) & 0.959 (2.553) \\
$(5,1)$ & 0.920 (5.819) & 0.953 (5.078) & 0.908 (4.913) & 0.925 (4.431) & 0.897 (6.587) & 0.943 (5.703) & 0.903 (5.619) & 0.921 (4.972) & 0.878 (8.945) & 0.957 (7.884) & 0.916 (8.018) & 0.942 (6.405) \\
$(5,2)$ & 0.886 (5.968) & 0.923 (5.192) & 0.881 (5.019) & 0.925 (4.518) & 0.889 (6.644) & 0.942 (5.768) & 0.926 (5.605) & 0.951 (4.974) & 0.868 (8.844) & 0.956 (7.872) & 0.926 (8.119) & 0.947 (6.435) \\
$(5,5)$ & 0.903 (5.894) & 0.937 (5.107) & 0.922 (4.940) & 0.958 (4.448) & 0.890 (6.619) & 0.947 (5.733) & 0.937 (5.610) & 0.955 (4.958) & 0.860 (9.202) & 0.952 (8.053) & 0.921 (8.215) & 0.954 (6.544) \\
\addlinespace[2pt]
\midrule
\addlinespace[2pt]
\multicolumn{13}{c}{\textbf{Panel: $n=30$}} \\
\addlinespace[1pt]
$(\alpha,\theta)$ & \multicolumn{4}{c}{20\% censoring} & \multicolumn{4}{c}{30\% censoring} & \multicolumn{4}{c}{50\% censoring} \\
\cmidrule(lr){2-5}\cmidrule(lr){6-9}\cmidrule(lr){10-13}
 & WLMA & B--LSE & B--MLE & A--MLE & WLMA & B--LSE & B--MLE & A--MLE & WLMA & B--LSE & B--MLE & A--MLE \\
\midrule
$(2,1)$ & 0.890 (1.912) & 0.938 (1.675) & 0.918 (1.541) & 0.948 (1.447) & 0.864 (2.129) & 0.938 (1.855) & 0.919 (1.726) & 0.949 (1.603) & 0.839 (2.817) & 0.933 (2.391) & 0.918 (2.342) & 0.947 (2.046) \\
$(2,2)$ & 0.903 (1.870) & 0.927 (1.646) & 0.939 (1.512) & 0.956 (1.421) & 0.883 (2.095) & 0.928 (1.819) & 0.938 (1.700) & 0.961 (1.579) & 0.842 (2.744) & 0.944 (2.334) & 0.926 (2.274) & 0.946 (1.997) \\
$(2,5)$ & 0.908 (1.874) & 0.945 (1.641) & 0.932 (1.510) & 0.954 (1.424) & 0.891 (2.086) & 0.927 (1.814) & 0.934 (1.704) & 0.949 (1.582) & 0.860 (2.756) & 0.927 (2.351) & 0.930 (2.324) & 0.956 (2.030) \\
$(5,1)$ & 0.893 (4.721) & 0.925 (4.135) & 0.916 (3.784) & 0.922 (3.553) & 0.863 (5.250) & 0.936 (4.553) & 0.898 (4.228) & 0.919 (3.929) & 0.844 (6.932) & 0.925 (5.871) & 0.921 (5.769) & 0.925 (5.054) \\
$(5,2)$ & 0.877 (4.708) & 0.927 (4.138) & 0.931 (3.800) & 0.948 (3.571) & 0.867 (5.312) & 0.929 (4.594) & 0.922 (4.287) & 0.940 (3.981) & 0.857 (6.861) & 0.936 (5.842) & 0.930 (5.738) & 0.956 (5.032) \\
$(5,5)$ & 0.900 (4.739) & 0.931 (4.149) & 0.935 (3.835) & 0.959 (3.601) & 0.878 (5.271) & 0.927 (4.573) & 0.944 (4.283) & 0.958 (3.976) & 0.845 (6.930) & 0.948 (5.862) & 0.940 (5.734) & 0.963 (5.036) \\
\addlinespace[2pt]
\midrule
\addlinespace[2pt]
\multicolumn{13}{c}{\textbf{Panel: $n=50$}} \\
\addlinespace[1pt]
$(\alpha,\theta)$ & \multicolumn{4}{c}{20\% censoring} & \multicolumn{4}{c}{30\% censoring} & \multicolumn{4}{c}{50\% censoring} \\
\cmidrule(lr){2-5}\cmidrule(lr){6-9}\cmidrule(lr){10-13}
 & WLMA & B--LSE & B--MLE & A--MLE & WLMA & B--LSE & B--MLE & A--MLE & WLMA & B--LSE & B--MLE & A--MLE \\
\midrule
$(2,1)$ & 0.858 (1.443) & 0.907 (1.299) & 0.935 (1.131) & 0.928 (1.096) & 0.849 (1.603) & 0.932 (1.421) & 0.933 (1.250) & 0.955 (1.204) & 0.814 (2.053) & 0.918 (1.762) & 0.939 (1.634) & 0.938 (1.520) \\
$(2,2)$ & 0.884 (1.441) & 0.928 (1.284) & 0.945 (1.126) & 0.954 (1.088) & 0.818 (1.615) & 0.924 (1.423) & 0.925 (1.256) & 0.947 (1.210) & 0.816 (2.036) & 0.922 (1.756) & 0.932 (1.625) & 0.942 (1.517) \\
$(2,5)$ & 0.874 (1.448) & 0.915 (1.301) & 0.923 (1.146) & 0.942 (1.107) & 0.851 (1.599) & 0.920 (1.414) & 0.936 (1.254) & 0.955 (1.202) & 0.807 (2.068) & 0.923 (1.780) & 0.933 (1.644) & 0.953 (1.531) \\
$(5,1)$ & 0.866 (3.601) & 0.914 (3.220) & 0.922 (2.827) & 0.934 (2.733) & 0.857 (3.987) & 0.936 (3.535) & 0.913 (3.121) & 0.938 (3.006) & 0.810 (5.138) & 0.937 (4.398) & 0.921 (4.065) & 0.938 (3.800) \\
$(5,2)$ & 0.883 (3.593) & 0.932 (3.200) & 0.941 (2.786) & 0.955 (2.705) & 0.849 (3.982) & 0.930 (3.528) & 0.932 (3.115) & 0.938 (2.999) & 0.815 (5.100) & 0.920 (4.369) & 0.932 (4.029) & 0.957 (3.761) \\
$(5,5)$ & 0.888 (3.592) & 0.927 (3.225) & 0.937 (2.829) & 0.962 (2.732) & 0.860 (3.981) & 0.927 (3.528) & 0.949 (3.116) & 0.960 (2.991) & 0.815 (5.112) & 0.938 (4.387) & 0.935 (4.058) & 0.954 (3.780) \\
\bottomrule
\end{tabular}
\par\vspace{3pt}
\begin{minipage}{0.98\linewidth}
\fontsize{6.5}{7.2}\selectfont
\emph{Note.} B--LSE denotes Bootstrap--LSE, B--MLE denotes Bootstrap--MLE, and A--MLE denotes asymptotic MLE.
\end{minipage}
\end{table}
\end{landscape}

\bigskip

Tables~\ref{tab:type1_alpha_balanced} and
\ref{tab:type1_alpha_transposed} summarize the empirical coverage probabilities
and average interval lengths for \(\alpha\). For all methods, increasing \(n\)
reduced interval length, while increasing the censoring proportion generally
widened the intervals. The KM- and HJ-based GLA results were very similar:
coverage was generally close to 0.95 under 20\% and 30\% censoring and declined
modestly under 50\% censoring, especially for \(n=20\). The differences between
KM and HJ were small relative to the effects of sample size and censoring, and
neither plotting estimator was uniformly better.

The comparison procedures displayed different coverage--length trade-offs.
WLMA generally produced shorter intervals with coverage below the nominal level
in a number of settings. Bootstrap--LSE often gave longer intervals than
Bootstrap--MLE and asymptotic MLE, but shorter intervals than GLA.
For example, at \(n=50\), \((\alpha,\theta)=(2,2)\), and 20\% censoring,
Bootstrap--MLE attained coverage 0.945 with average length 1.126, compared with
0.964 (1.646) for GLA--KM and 0.966 (1.652) for GLA--HJ. Asymptotic MLE
attained coverage 0.954 with average length 1.088 in this setting.

\begin{table}[htbp]
  \centering
  \caption{Average interval length and coverage probability for the scale parameter $\theta$ using the proposed GLA--KM and GLA--HJ intervals under Type-I censoring (95\% confidence intervals; 2,000 Monte Carlo replications; 1,000 GPQ draws)}
  \label{tab:type1_theta_balanced}
  \small
  \resizebox{\textwidth}{!}{%
  \begin{tabular}{|c|c|c|c|c|c|c|c|c|}
    \hline
    \multirow{2}{*}{$n$} & \multirow{2}{*}{Parameter setting} & \multirow{2}{*}{Method} & \multicolumn{2}{c|}{20\% censoring} & \multicolumn{2}{c|}{30\% censoring} & \multicolumn{2}{c|}{50\% censoring} \\
    \cline{4-9}
    & & & Length & Coverage & Length & Coverage & Length & Coverage \\
    \hline
    \multirow{12}{*}{20} & \multirow{2}{*}{$\alpha=2,\ \theta=1$} & KM & 5.541 & 0.970 & 5.252 & 0.971 & 3.899 & 0.930 \\
    \cline{3-9}
     &  & HJ & 5.816 & 0.967 & 5.519 & 0.974 & 4.079 & 0.938 \\
    \cline{2-9}
     & \multirow{2}{*}{$\alpha=2,\ \theta=2$} & KM & 12.247 & 0.969 & 10.816 & 0.966 & 11.041 & 0.949 \\
    \cline{3-9}
     &  & HJ & 11.448 & 0.978 & 10.833 & 0.963 & 12.020 & 0.942 \\
    \cline{2-9}
     & \multirow{2}{*}{$\alpha=2,\ \theta=5$} & KM & 27.924 & 0.971 & 27.680 & 0.964 & 20.072 & 0.928 \\
    \cline{3-9}
     &  & HJ & 29.316 & 0.982 & 27.062 & 0.962 & 19.039 & 0.951 \\
    \cline{2-9}
     & \multirow{2}{*}{$\alpha=5,\ \theta=1$} & KM & 1.091 & 0.971 & 1.052 & 0.966 & 0.933 & 0.927 \\
    \cline{3-9}
     &  & HJ & 1.085 & 0.978 & 1.072 & 0.963 & 0.903 & 0.926 \\
    \cline{2-9}
     & \multirow{2}{*}{$\alpha=5,\ \theta=2$} & KM & 2.105 & 0.970 & 2.128 & 0.960 & 1.937 & 0.942 \\
    \cline{3-9}
     &  & HJ & 2.160 & 0.967 & 2.129 & 0.969 & 1.904 & 0.935 \\
    \cline{2-9}
     & \multirow{2}{*}{$\alpha=5,\ \theta=5$} & KM & 5.456 & 0.964 & 5.297 & 0.966 & 4.488 & 0.942 \\
    \cline{3-9}
     &  & HJ & 5.410 & 0.972 & 5.572 & 0.967 & 4.595 & 0.934 \\
    \hline
    \multirow{12}{*}{30} & \multirow{2}{*}{$\alpha=2,\ \theta=1$} & KM & 5.162 & 0.981 & 5.198 & 0.971 & 3.735 & 0.950 \\
    \cline{3-9}
     &  & HJ & 5.316 & 0.979 & 5.042 & 0.970 & 3.951 & 0.949 \\
    \cline{2-9}
     & \multirow{2}{*}{$\alpha=2,\ \theta=2$} & KM & 11.166 & 0.979 & 11.228 & 0.962 & 7.912 & 0.953 \\
    \cline{3-9}
     &  & HJ & 10.635 & 0.971 & 10.408 & 0.969 & 7.995 & 0.937 \\
    \cline{2-9}
     & \multirow{2}{*}{$\alpha=2,\ \theta=5$} & KM & 26.187 & 0.978 & 26.510 & 0.962 & 19.505 & 0.950 \\
    \cline{3-9}
     &  & HJ & 28.953 & 0.964 & 26.405 & 0.972 & 20.154 & 0.945 \\
    \cline{2-9}
     & \multirow{2}{*}{$\alpha=5,\ \theta=1$} & KM & 1.074 & 0.971 & 1.084 & 0.965 & 0.954 & 0.950 \\
    \cline{3-9}
     &  & HJ & 1.026 & 0.980 & 1.082 & 0.964 & 0.939 & 0.946 \\
    \cline{2-9}
     & \multirow{2}{*}{$\alpha=5,\ \theta=2$} & KM & 2.105 & 0.962 & 2.090 & 0.962 & 1.848 & 0.938 \\
    \cline{3-9}
     &  & HJ & 2.122 & 0.970 & 2.134 & 0.972 & 1.919 & 0.942 \\
    \cline{2-9}
     & \multirow{2}{*}{$\alpha=5,\ \theta=5$} & KM & 5.214 & 0.971 & 5.299 & 0.965 & 4.727 & 0.956 \\
    \cline{3-9}
     &  & HJ & 5.160 & 0.972 & 5.264 & 0.965 & 4.662 & 0.941 \\
    \hline
    \multirow{12}{*}{50} & \multirow{2}{*}{$\alpha=2,\ \theta=1$} & KM & 4.966 & 0.972 & 4.878 & 0.966 & 4.012 & 0.953 \\
    \cline{3-9}
     &  & HJ & 4.911 & 0.973 & 5.077 & 0.971 & 4.055 & 0.958 \\
    \cline{2-9}
     & \multirow{2}{*}{$\alpha=2,\ \theta=2$} & KM & 9.884 & 0.965 & 9.875 & 0.970 & 8.367 & 0.954 \\
    \cline{3-9}
     &  & HJ & 9.523 & 0.966 & 9.598 & 0.968 & 8.505 & 0.964 \\
    \cline{2-9}
     & \multirow{2}{*}{$\alpha=2,\ \theta=5$} & KM & 24.068 & 0.965 & 24.780 & 0.965 & 21.020 & 0.964 \\
    \cline{3-9}
     &  & HJ & 25.909 & 0.979 & 26.847 & 0.975 & 20.256 & 0.949 \\
    \cline{2-9}
     & \multirow{2}{*}{$\alpha=5,\ \theta=1$} & KM & 0.959 & 0.972 & 1.003 & 0.971 & 0.964 & 0.965 \\
    \cline{3-9}
     &  & HJ & 0.944 & 0.975 & 1.035 & 0.976 & 0.964 & 0.961 \\
    \cline{2-9}
     & \multirow{2}{*}{$\alpha=5,\ \theta=2$} & KM & 1.905 & 0.980 & 1.992 & 0.974 & 1.936 & 0.963 \\
    \cline{3-9}
     &  & HJ & 1.847 & 0.972 & 1.970 & 0.966 & 1.890 & 0.950 \\
    \cline{2-9}
     & \multirow{2}{*}{$\alpha=5,\ \theta=5$} & KM & 4.676 & 0.975 & 5.036 & 0.977 & 4.813 & 0.964 \\
    \cline{3-9}
     &  & HJ & 4.849 & 0.968 & 5.108 & 0.968 & 4.835 & 0.969 \\
    \hline
  \end{tabular}%
  }

  \begin{minipage}{0.98\textwidth}
  \footnotesize
  \emph{Note.} The fixed censoring time \(C\) was determined by the target censoring proportion, as described in the simulation design. KM denotes the Kaplan Meier plotting estimator and HJ denotes the Herd--Johnson plotting estimator.
  \end{minipage}
\end{table}


\begin{landscape}
\begin{table}[p]
\centering
\caption{Coverage probability and average interval length for the scale parameter $\theta$ using the comparison methods under Type-I censoring (95\% confidence intervals)}
\label{tab:type1_theta_transposed}
\fontsize{7.2}{8.1}\selectfont
\setlength{\tabcolsep}{1.8pt}
\renewcommand{\arraystretch}{0.94}
\textit{Entries are coverage probability (average interval length).}\par\vspace{3pt}
\begin{tabular}{@{}c*{12}{c}@{}}
\toprule
\multicolumn{13}{c}{\textbf{Panel: $n=20$}} \\
\addlinespace[1pt]
$(\alpha,\theta)$ & \multicolumn{4}{c}{20\% censoring} & \multicolumn{4}{c}{30\% censoring} & \multicolumn{4}{c}{50\% censoring} \\
\cmidrule(lr){2-5}\cmidrule(lr){6-9}\cmidrule(lr){10-13}
 & WLMA & B--LSE & B--MLE & A--MLE & WLMA & B--LSE & B--MLE & A--MLE & WLMA & B--LSE & B--MLE & A--MLE \\
\midrule
$(2,1)$ & 0.663 (0.430) & 0.945 (0.940) & 0.923 (0.537) & 0.931 (0.504) & 0.382 (0.411) & 0.949 (1.551) & 0.937 (0.635) & 0.942 (0.557) & 0.107 (0.410) & 0.934 (39.969) & 0.929 (1.279) & 0.904 (0.867) \\
$(2,2)$ & 0.664 (0.852) & 0.940 (1.966) & 0.933 (1.061) & 0.939 (0.995) & 0.351 (0.815) & 0.949 (3.517) & 0.927 (1.259) & 0.933 (1.109) & 0.082 (0.802) & 0.951 (62.060) & 0.940 (2.295) & 0.904 (1.629) \\
$(2,5)$ & 0.678 (2.149) & 0.961 (4.597) & 0.923 (2.645) & 0.933 (2.482) & 0.373 (2.069) & 0.961 (7.274) & 0.929 (3.248) & 0.948 (2.852) & 0.104 (2.031) & 0.945 (71.368) & 0.936 (5.493) & 0.893 (3.953) \\
$(5,1)$ & 0.661 (0.192) & 0.948 (0.317) & 0.901 (0.207) & 0.907 (0.198) & 0.403 (0.197) & 0.944 (0.423) & 0.899 (0.236) & 0.922 (0.217) & 0.112 (0.222) & 0.947 (1.664) & 0.923 (0.380) & 0.908 (0.313) \\
$(5,2)$ & 0.656 (0.382) & 0.946 (0.632) & 0.913 (0.411) & 0.931 (0.394) & 0.390 (0.389) & 0.942 (0.820) & 0.920 (0.471) & 0.938 (0.431) & 0.108 (0.446) & 0.944 (2.312) & 0.934 (0.761) & 0.927 (0.626) \\
$(5,5)$ & 0.647 (0.955) & 0.942 (1.555) & 0.924 (1.032) & 0.937 (0.988) & 0.370 (0.992) & 0.943 (2.244) & 0.931 (1.195) & 0.950 (1.102) & 0.111 (1.073) & 0.932 (4.996) & 0.943 (1.823) & 0.919 (1.511) \\
\addlinespace[2pt]
\midrule
\addlinespace[2pt]
\multicolumn{13}{c}{\textbf{Panel: $n=30$}} \\
\addlinespace[1pt]
$(\alpha,\theta)$ & \multicolumn{4}{c}{20\% censoring} & \multicolumn{4}{c}{30\% censoring} & \multicolumn{4}{c}{50\% censoring} \\
\cmidrule(lr){2-5}\cmidrule(lr){6-9}\cmidrule(lr){10-13}
 & WLMA & B--LSE & B--MLE & A--MLE & WLMA & B--LSE & B--MLE & A--MLE & WLMA & B--LSE & B--MLE & A--MLE \\
\midrule
$(2,1)$ & 0.467 (0.331) & 0.958 (0.633) & 0.924 (0.413) & 0.937 (0.401) & 0.151 (0.320) & 0.958 (0.901) & 0.939 (0.472) & 0.952 (0.440) & 0.015 (0.303) & 0.963 (2.998) & 0.926 (0.815) & 0.913 (0.654) \\
$(2,2)$ & 0.458 (0.674) & 0.954 (1.295) & 0.938 (0.833) & 0.940 (0.807) & 0.145 (0.648) & 0.943 (1.871) & 0.928 (0.980) & 0.940 (0.907) & 0.007 (0.604) & 0.953 (5.731) & 0.943 (1.595) & 0.926 (1.284) \\
$(2,5)$ & 0.442 (1.661) & 0.943 (3.110) & 0.933 (2.071) & 0.933 (1.999) & 0.171 (1.641) & 0.956 (4.775) & 0.932 (2.439) & 0.943 (2.262) & 0.013 (1.545) & 0.956 (17.905) & 0.928 (3.921) & 0.915 (3.192) \\
$(5,1)$ & 0.459 (0.152) & 0.950 (0.241) & 0.903 (0.167) & 0.921 (0.163) & 0.149 (0.153) & 0.961 (0.298) & 0.904 (0.184) & 0.923 (0.176) & 0.012 (0.166) & 0.958 (0.604) & 0.921 (0.274) & 0.916 (0.243) \\
$(5,2)$ & 0.476 (0.305) & 0.957 (0.483) & 0.931 (0.331) & 0.950 (0.324) & 0.174 (0.307) & 0.953 (0.619) & 0.937 (0.368) & 0.956 (0.350) & 0.017 (0.329) & 0.944 (1.189) & 0.940 (0.533) & 0.933 (0.474) \\
$(5,5)$ & 0.472 (0.762) & 0.945 (1.211) & 0.946 (0.829) & 0.955 (0.810) & 0.140 (0.768) & 0.954 (1.531) & 0.937 (0.915) & 0.949 (0.874) & 0.016 (0.824) & 0.950 (2.959) & 0.942 (1.355) & 0.938 (1.208) \\
\addlinespace[2pt]
\midrule
\addlinespace[2pt]
\multicolumn{13}{c}{\textbf{Panel: $n=50$}} \\
\addlinespace[1pt]
$(\alpha,\theta)$ & \multicolumn{4}{c}{20\% censoring} & \multicolumn{4}{c}{30\% censoring} & \multicolumn{4}{c}{50\% censoring} \\
\cmidrule(lr){2-5}\cmidrule(lr){6-9}\cmidrule(lr){10-13}
 & WLMA & B--LSE & B--MLE & A--MLE & WLMA & B--LSE & B--MLE & A--MLE & WLMA & B--LSE & B--MLE & A--MLE \\
\midrule
$(2,1)$ & 0.209 (0.254) & 0.955 (0.452) & 0.927 (0.315) & 0.942 (0.310) & 0.017 (0.240) & 0.942 (0.577) & 0.939 (0.352) & 0.938 (0.339) & 0.000 (0.225) & 0.957 (1.253) & 0.940 (0.526) & 0.927 (0.474) \\
$(2,2)$ & 0.188 (0.506) & 0.950 (0.908) & 0.942 (0.636) & 0.948 (0.628) & 0.015 (0.478) & 0.952 (1.166) & 0.939 (0.699) & 0.955 (0.675) & 0.000 (0.452) & 0.961 (2.504) & 0.936 (1.042) & 0.937 (0.942) \\
$(2,5)$ & 0.221 (1.267) & 0.953 (2.289) & 0.929 (1.581) & 0.938 (1.552) & 0.013 (1.206) & 0.953 (2.947) & 0.937 (1.769) & 0.949 (1.707) & 0.000 (1.122) & 0.943 (6.322) & 0.945 (2.596) & 0.938 (2.339) \\
$(5,1)$ & 0.198 (0.115) & 0.957 (0.175) & 0.919 (0.126) & 0.917 (0.125) & 0.018 (0.116) & 0.947 (0.213) & 0.920 (0.137) & 0.933 (0.134) & 0.000 (0.122) & 0.962 (0.375) & 0.939 (0.192) & 0.929 (0.181) \\
$(5,2)$ & 0.168 (0.230) & 0.954 (0.346) & 0.936 (0.254) & 0.942 (0.252) & 0.016 (0.231) & 0.959 (0.423) & 0.931 (0.274) & 0.936 (0.268) & 0.000 (0.247) & 0.954 (0.769) & 0.942 (0.394) & 0.942 (0.369) \\
$(5,5)$ & 0.203 (0.574) & 0.964 (0.862) & 0.944 (0.625) & 0.950 (0.619) & 0.015 (0.581) & 0.962 (1.070) & 0.942 (0.688) & 0.954 (0.674) & 0.000 (0.612) & 0.959 (1.884) & 0.937 (0.966) & 0.932 (0.913) \\
\bottomrule
\end{tabular}
\par\vspace{3pt}
\begin{minipage}{0.98\linewidth}
\fontsize{6.5}{7.2}\selectfont
\emph{Note.} B--LSE denotes Bootstrap--LSE, B--MLE denotes Bootstrap--MLE, and A--MLE denotes asymptotic MLE.
\end{minipage}
\end{table}
\end{landscape}

Tables~\ref{tab:type1_theta_balanced} and
\ref{tab:type1_theta_transposed} summarize the empirical coverage probabilities
and average interval lengths for \(\theta\). The KM- and HJ-based GLA intervals
again had nearly identical coverage and length. Their coverage remained close
to the nominal level across most settings, including under 50\% censoring, and
their lengths generally decreased with \(n\). Thus, the choice between KM and
HJ had much less influence than sample size, censoring, and the parameter
configuration.

The effect of censoring differed among the comparison methods. WLMA generally
produced relatively short intervals, with coverage farther from the nominal
level in several heavier-censoring settings. \citet{Xiang2015} provided a
useful assessment of one-sided confidence limits, including lower limits for
the scale parameter, which is different from the assessment of two-sided
intervals in the present study.
Bootstrap--LSE usually retained
near-nominal coverage, although a few heavy-censoring scenarios produced very
large average lengths. At \(n=20\), \(\alpha=2\), and 50\% censoring,
few observed failures can lead to extreme LSE-based bootstrap scale estimates
and very long intervals. These occasional intervals make the mean length
sensitive to the Monte Carlo sample.
Bootstrap--MLE generally attained coverage between about
0.90 and 0.95, while asymptotic MLE often gave the shortest intervals.
Near-nominal GLA coverage could be accompanied by substantially greater length.
For example, at \(n=50\), \((\alpha,\theta)=(2,2)\), and 20\% censoring,
Bootstrap--LSE attained coverage 0.950 with average length 0.908, compared with
0.965 (9.884) for GLA--KM and 0.966 (9.523) for GLA--HJ. Thus, in this
configuration, Bootstrap--LSE provided coverage closer to the nominal level
with much shorter intervals.

\begin{table}[htbp]
  \centering
  \caption{Average interval length and coverage probability for $S(1.2)$ using the proposed GLA--KM and GLA--HJ intervals under Type-I censoring (95\% confidence intervals; 2,000 Monte Carlo replications; 1,000 GPQ draws)}
  \label{tab:type1_survival_t1_2}
  \small
  \resizebox{\textwidth}{!}{%
  \begin{tabular}{|c|c|c|c|c|c|c|c|c|}
    \hline
    \multirow{2}{*}{$n$} & \multirow{2}{*}{Parameter setting} & \multirow{2}{*}{Method} & \multicolumn{2}{c|}{20\% censoring} & \multicolumn{2}{c|}{30\% censoring} & \multicolumn{2}{c|}{50\% censoring} \\
    \cline{4-9}
    & & & Length & Coverage & Length & Coverage & Length & Coverage \\
    \hline
    \multirow{12}{*}{20} & \multirow{2}{*}{$\alpha=2,\ \theta=1$} & KM & 0.773 & 0.971 & 0.779 & 0.959 & 0.728 & 0.939 \\
    \cline{3-9}
     &  & HJ & 0.779 & 0.971 & 0.772 & 0.955 & 0.712 & 0.928 \\
    \cline{2-9}
     & \multirow{2}{*}{$\alpha=2,\ \theta=2$} & KM & 0.466 & 0.960 & 0.534 & 0.955 & 0.719 & 0.946 \\
    \cline{3-9}
     &  & HJ & 0.460 & 0.954 & 0.533 & 0.968 & 0.716 & 0.947 \\
    \cline{2-9}
     & \multirow{2}{*}{$\alpha=2,\ \theta=5$} & KM & 0.147 & 0.943 & 0.165 & 0.941 & 0.235 & 0.962 \\
    \cline{3-9}
     &  & HJ & 0.152 & 0.954 & 0.162 & 0.937 & 0.240 & 0.948 \\
    \cline{2-9}
     & \multirow{2}{*}{$\alpha=5,\ \theta=1$} & KM & 0.702 & 0.964 & 0.707 & 0.969 & 0.614 & 0.930 \\
    \cline{3-9}
     &  & HJ & 0.725 & 0.974 & 0.716 & 0.963 & 0.629 & 0.940 \\
    \cline{2-9}
     & \multirow{2}{*}{$\alpha=5,\ \theta=2$} & KM & 0.177 & 0.938 & 0.206 & 0.952 & 0.289 & 0.954 \\
    \cline{3-9}
     &  & HJ & 0.181 & 0.946 & 0.202 & 0.947 & 0.284 & 0.948 \\
    \cline{2-9}
     & \multirow{2}{*}{$\alpha=5,\ \theta=5$} & KM & 0.012 & 0.939 & 0.013 & 0.936 & 0.022 & 0.925 \\
    \cline{3-9}
     &  & HJ & 0.011 & 0.943 & 0.013 & 0.919 & 0.021 & 0.896 \\
    \hline
    \multirow{12}{*}{30} & \multirow{2}{*}{$\alpha=2,\ \theta=1$} & KM & 0.751 & 0.964 & 0.781 & 0.965 & 0.741 & 0.938 \\
    \cline{3-9}
     &  & HJ & 0.747 & 0.969 & 0.778 & 0.969 & 0.749 & 0.950 \\
    \cline{2-9}
     & \multirow{2}{*}{$\alpha=2,\ \theta=2$} & KM & 0.417 & 0.959 & 0.494 & 0.966 & 0.687 & 0.960 \\
    \cline{3-9}
     &  & HJ & 0.421 & 0.963 & 0.493 & 0.971 & 0.696 & 0.962 \\
    \cline{2-9}
     & \multirow{2}{*}{$\alpha=2,\ \theta=5$} & KM & 0.121 & 0.944 & 0.136 & 0.943 & 0.193 & 0.961 \\
    \cline{3-9}
     &  & HJ & 0.124 & 0.938 & 0.136 & 0.945 & 0.191 & 0.957 \\
    \cline{2-9}
     & \multirow{2}{*}{$\alpha=5,\ \theta=1$} & KM & 0.717 & 0.976 & 0.730 & 0.972 & 0.663 & 0.945 \\
    \cline{3-9}
     &  & HJ & 0.705 & 0.970 & 0.710 & 0.967 & 0.647 & 0.941 \\
    \cline{2-9}
     & \multirow{2}{*}{$\alpha=5,\ \theta=2$} & KM & 0.151 & 0.944 & 0.169 & 0.961 & 0.241 & 0.963 \\
    \cline{3-9}
     &  & HJ & 0.148 & 0.940 & 0.166 & 0.960 & 0.234 & 0.969 \\
    \cline{2-9}
     & \multirow{2}{*}{$\alpha=5,\ \theta=5$} & KM & 0.008 & 0.955 & 0.008 & 0.931 & 0.012 & 0.909 \\
    \cline{3-9}
     &  & HJ & 0.009 & 0.955 & 0.008 & 0.935 & 0.012 & 0.917 \\
    \hline
    \multirow{12}{*}{50} & \multirow{2}{*}{$\alpha=2,\ \theta=1$} & KM & 0.718 & 0.970 & 0.786 & 0.973 & 0.785 & 0.956 \\
    \cline{3-9}
     &  & HJ & 0.710 & 0.967 & 0.784 & 0.971 & 0.773 & 0.950 \\
    \cline{2-9}
     & \multirow{2}{*}{$\alpha=2,\ \theta=2$} & KM & 0.370 & 0.955 & 0.449 & 0.974 & 0.644 & 0.959 \\
    \cline{3-9}
     &  & HJ & 0.367 & 0.969 & 0.444 & 0.971 & 0.651 & 0.960 \\
    \cline{2-9}
     & \multirow{2}{*}{$\alpha=2,\ \theta=5$} & KM & 0.097 & 0.927 & 0.108 & 0.944 & 0.152 & 0.974 \\
    \cline{3-9}
     &  & HJ & 0.097 & 0.933 & 0.108 & 0.939 & 0.154 & 0.972 \\
    \cline{2-9}
     & \multirow{2}{*}{$\alpha=5,\ \theta=1$} & KM & 0.673 & 0.968 & 0.721 & 0.972 & 0.689 & 0.952 \\
    \cline{3-9}
     &  & HJ & 0.670 & 0.961 & 0.721 & 0.967 & 0.705 & 0.964 \\
    \cline{2-9}
     & \multirow{2}{*}{$\alpha=5,\ \theta=2$} & KM & 0.122 & 0.956 & 0.136 & 0.951 & 0.196 & 0.977 \\
    \cline{3-9}
     &  & HJ & 0.121 & 0.929 & 0.137 & 0.951 & 0.194 & 0.972 \\
    \cline{2-9}
     & \multirow{2}{*}{$\alpha=5,\ \theta=5$} & KM & 0.005 & 0.962 & 0.005 & 0.931 & 0.007 & 0.924 \\
    \cline{3-9}
     &  & HJ & 0.005 & 0.945 & 0.005 & 0.939 & 0.007 & 0.928 \\
    \hline
  \end{tabular}%
  }

  \begin{minipage}{0.98\textwidth}
  \footnotesize
  \emph{Note.} The fixed censoring time \(C\) was determined by the target censoring proportion, as described in the simulation design. KM denotes the Kaplan--Meier plotting estimator and HJ denotes the Herd--Johnson plotting estimator.
  \end{minipage}
\end{table}

\bigskip

\begin{landscape}
\begin{table}[p]
\centering
\caption{Coverage probability and average interval length for $S(1.2)$ using the comparison methods under Type-I censoring (95\% confidence intervals; 2,000 Monte Carlo replications; 1,000 GPQ or bootstrap draws where applicable)}
\label{tab:type1_s12_transposed}
\fontsize{7.2}{8.1}\selectfont
\setlength{\tabcolsep}{1.8pt}
\renewcommand{\arraystretch}{0.94}
\textit{Entries are coverage probability (average interval length).}\par\vspace{3pt}
\begin{tabular}{@{}c*{12}{c}@{}}
\toprule
\multicolumn{13}{c}{\textbf{Panel: $n=20$}} \\
\addlinespace[1pt]
$(\alpha,\theta)$ & \multicolumn{4}{c}{20\% censoring} & \multicolumn{4}{c}{30\% censoring} & \multicolumn{4}{c}{50\% censoring} \\
\cmidrule(lr){2-5}\cmidrule(lr){6-9}\cmidrule(lr){10-13}
 & WLMA & B--LSE & B--MLE & A--MLE & WLMA & B--LSE & B--MLE & A--MLE & WLMA & B--LSE & B--MLE & A--MLE \\
\midrule
$(2,1)$ & 0.687 (0.255) & 0.957 (0.378) & 0.908 (0.326) & 0.895 (0.331) & 0.417 (0.212) & 0.951 (0.427) & 0.931 (0.359) & 0.911 (0.369) & 0.201 (0.147) & 0.951 (0.542) & 0.932 (0.469) & 0.851 (0.460) \\
$(2,2)$ & 0.803 (0.374) & 0.933 (0.326) & 0.926 (0.321) & 0.926 (0.326) & 0.615 (0.401) & 0.921 (0.331) & 0.918 (0.323) & 0.902 (0.325) & 0.180 (0.453) & 0.934 (0.363) & 0.924 (0.334) & 0.919 (0.331) \\
$(2,5)$ & 0.911 (0.197) & 0.914 (0.188) & 0.926 (0.130) & 0.872 (0.127) & 0.893 (0.227) & 0.934 (0.188) & 0.908 (0.133) & 0.852 (0.128) & 0.826 (0.312) & 0.935 (0.190) & 0.917 (0.136) & 0.856 (0.132) \\
$(5,1)$ & 0.720 (0.129) & 0.959 (0.318) & 0.897 (0.225) & 0.825 (0.204) & 0.568 (0.105) & 0.962 (0.370) & 0.897 (0.261) & 0.811 (0.225) & 0.355 (0.076) & 0.959 (0.477) & 0.926 (0.366) & 0.756 (0.292) \\
$(5,2)$ & 0.900 (0.223) & 0.921 (0.205) & 0.891 (0.152) & 0.850 (0.152) & 0.885 (0.255) & 0.936 (0.208) & 0.916 (0.156) & 0.874 (0.155) & 0.774 (0.354) & 0.943 (0.213) & 0.906 (0.158) & 0.866 (0.158) \\
$(5,5)$ & 0.909 (0.019) & 0.932 (0.032) & 0.916 (0.010) & 0.765 (0.005) & 0.905 (0.025) & 0.942 (0.036) & 0.933 (0.011) & 0.767 (0.006) & 0.898 (0.042) & 0.944 (0.040) & 0.905 (0.013) & 0.719 (0.007) \\
\addlinespace[2pt]
\midrule
\addlinespace[2pt]
\multicolumn{13}{c}{\textbf{Panel: $n=30$}} \\
\addlinespace[1pt]
$(\alpha,\theta)$ & \multicolumn{4}{c}{20\% censoring} & \multicolumn{4}{c}{30\% censoring} & \multicolumn{4}{c}{50\% censoring} \\
\cmidrule(lr){2-5}\cmidrule(lr){6-9}\cmidrule(lr){10-13}
 & WLMA & B--LSE & B--MLE & A--MLE & WLMA & B--LSE & B--MLE & A--MLE & WLMA & B--LSE & B--MLE & A--MLE \\
\midrule
$(2,1)$ & 0.466 (0.210) & 0.965 (0.323) & 0.914 (0.270) & 0.902 (0.274) & 0.202 (0.174) & 0.960 (0.374) & 0.938 (0.300) & 0.922 (0.308) & 0.049 (0.101) & 0.968 (0.500) & 0.922 (0.407) & 0.883 (0.410) \\
$(2,2)$ & 0.733 (0.313) & 0.933 (0.267) & 0.931 (0.266) & 0.931 (0.269) & 0.485 (0.336) & 0.944 (0.270) & 0.938 (0.268) & 0.938 (0.269) & 0.068 (0.378) & 0.935 (0.292) & 0.921 (0.275) & 0.923 (0.274) \\
$(2,5)$ & 0.933 (0.158) & 0.930 (0.151) & 0.937 (0.110) & 0.902 (0.112) & 0.881 (0.186) & 0.910 (0.153) & 0.936 (0.112) & 0.902 (0.113) & 0.775 (0.257) & 0.932 (0.153) & 0.918 (0.114) & 0.877 (0.115) \\
$(5,1)$ & 0.579 (0.098) & 0.964 (0.277) & 0.910 (0.193) & 0.866 (0.185) & 0.343 (0.070) & 0.962 (0.317) & 0.893 (0.216) & 0.830 (0.198) & 0.181 (0.042) & 0.957 (0.435) & 0.924 (0.309) & 0.810 (0.256) \\
$(5,2)$ & 0.903 (0.186) & 0.926 (0.170) & 0.936 (0.131) & 0.907 (0.135) & 0.875 (0.210) & 0.929 (0.169) & 0.911 (0.131) & 0.882 (0.135) & 0.743 (0.291) & 0.934 (0.172) & 0.936 (0.135) & 0.903 (0.140) \\
$(5,5)$ & 0.924 (0.011) & 0.933 (0.020) & 0.930 (0.007) & 0.789 (0.004) & 0.910 (0.014) & 0.924 (0.022) & 0.935 (0.007) & 0.803 (0.004) & 0.906 (0.024) & 0.945 (0.027) & 0.932 (0.009) & 0.771 (0.005) \\
\addlinespace[2pt]
\midrule
\addlinespace[2pt]
\multicolumn{13}{c}{\textbf{Panel: $n=50$}} \\
\addlinespace[1pt]
$(\alpha,\theta)$ & \multicolumn{4}{c}{20\% censoring} & \multicolumn{4}{c}{30\% censoring} & \multicolumn{4}{c}{50\% censoring} \\
\cmidrule(lr){2-5}\cmidrule(lr){6-9}\cmidrule(lr){10-13}
 & WLMA & B--LSE & B--MLE & A--MLE & WLMA & B--LSE & B--MLE & A--MLE & WLMA & B--LSE & B--MLE & A--MLE \\
\midrule
$(2,1)$ & 0.227 (0.169) & 0.948 (0.263) & 0.925 (0.213) & 0.918 (0.215) & 0.041 (0.135) & 0.949 (0.307) & 0.936 (0.237) & 0.923 (0.241) & 0.004 (0.069) & 0.960 (0.431) & 0.939 (0.327) & 0.918 (0.338) \\
$(2,2)$ & 0.608 (0.247) & 0.935 (0.207) & 0.929 (0.207) & 0.937 (0.210) & 0.294 (0.264) & 0.928 (0.208) & 0.931 (0.208) & 0.929 (0.210) & 0.010 (0.300) & 0.946 (0.223) & 0.931 (0.213) & 0.931 (0.213) \\
$(2,5)$ & 0.903 (0.120) & 0.919 (0.113) & 0.920 (0.084) & 0.891 (0.087) & 0.876 (0.142) & 0.927 (0.117) & 0.938 (0.089) & 0.902 (0.092) & 0.711 (0.201) & 0.922 (0.116) & 0.922 (0.092) & 0.891 (0.095) \\
$(5,1)$ & 0.377 (0.072) & 0.953 (0.228) & 0.920 (0.152) & 0.896 (0.154) & 0.157 (0.048) & 0.945 (0.268) & 0.914 (0.173) & 0.869 (0.169) & 0.035 (0.021) & 0.953 (0.376) & 0.934 (0.244) & 0.850 (0.214) \\
$(5,2)$ & 0.922 (0.147) & 0.935 (0.132) & 0.954 (0.105) & 0.931 (0.109) & 0.838 (0.166) & 0.917 (0.132) & 0.931 (0.106) & 0.905 (0.110) & 0.614 (0.237) & 0.925 (0.132) & 0.930 (0.111) & 0.910 (0.115) \\
$(5,5)$ & 0.930 (0.006) & 0.923 (0.012) & 0.939 (0.004) & 0.849 (0.003) & 0.916 (0.008) & 0.930 (0.013) & 0.943 (0.005) & 0.860 (0.003) & 0.925 (0.013) & 0.937 (0.017) & 0.938 (0.006) & 0.818 (0.004) \\
\bottomrule
\end{tabular}
\par\vspace{3pt}
\begin{minipage}{0.98\linewidth}
\fontsize{6.5}{7.2}\selectfont
\emph{Note.} B--LSE denotes Bootstrap--LSE, B--MLE denotes Bootstrap--MLE, and A--MLE denotes asymptotic MLE.
\end{minipage}
\end{table}
\end{landscape}

\bigskip

\begin{table}[htbp]
  \centering
  \caption{Average interval length and coverage probability for $S(1.5)$ using the proposed GLA--KM and GLA--HJ intervals under Type-I censoring (95\% confidence intervals; 2,000 Monte Carlo replications; 1,000 GPQ draws)}
  \label{tab:type1_survival_t1_5}
  \small
  \resizebox{\textwidth}{!}{%
  \begin{tabular}{|c|c|c|c|c|c|c|c|c|}
    \hline
    \multirow{2}{*}{$n$} & \multirow{2}{*}{Parameter setting} & \multirow{2}{*}{Method} & \multicolumn{2}{c|}{20\% censoring} & \multicolumn{2}{c|}{30\% censoring} & \multicolumn{2}{c|}{50\% censoring} \\
    \cline{4-9}
    & & & Length & Coverage & Length & Coverage & Length & Coverage \\
    \hline
    \multirow{12}{*}{20} & \multirow{2}{*}{$\alpha=2,\ \theta=1$} & KM & 0.730 & 0.972 & 0.719 & 0.961 & 0.645 & 0.939 \\
    \cline{3-9}
     &  & HJ & 0.741 & 0.970 & 0.711 & 0.954 & 0.625 & 0.933 \\
    \cline{2-9}
     & \multirow{2}{*}{$\alpha=2,\ \theta=2$} & KM & 0.598 & 0.969 & 0.681 & 0.964 & 0.823 & 0.945 \\
    \cline{3-9}
     &  & HJ & 0.594 & 0.974 & 0.680 & 0.967 & 0.816 & 0.934 \\
    \cline{2-9}
     & \multirow{2}{*}{$\alpha=2,\ \theta=5$} & KM & 0.195 & 0.945 & 0.220 & 0.951 & 0.313 & 0.959 \\
    \cline{3-9}
     &  & HJ & 0.200 & 0.955 & 0.217 & 0.950 & 0.318 & 0.957 \\
    \cline{2-9}
     & \multirow{2}{*}{$\alpha=5,\ \theta=1$} & KM & 0.522 & 0.963 & 0.515 & 0.964 & 0.390 & 0.920 \\
    \cline{3-9}
     &  & HJ & 0.553 & 0.972 & 0.522 & 0.962 & 0.412 & 0.933 \\
    \cline{2-9}
     & \multirow{2}{*}{$\alpha=5,\ \theta=2$} & KM & 0.358 & 0.946 & 0.419 & 0.973 & 0.577 & 0.959 \\
    \cline{3-9}
     &  & HJ & 0.361 & 0.943 & 0.413 & 0.971 & 0.574 & 0.951 \\
    \cline{2-9}
     & \multirow{2}{*}{$\alpha=5,\ \theta=5$} & KM & 0.022 & 0.940 & 0.024 & 0.933 & 0.039 & 0.928 \\
    \cline{3-9}
     &  & HJ & 0.021 & 0.941 & 0.025 & 0.924 & 0.037 & 0.905 \\
    \hline
    \multirow{12}{*}{30} & \multirow{2}{*}{$\alpha=2,\ \theta=1$} & KM & 0.714 & 0.963 & 0.727 & 0.964 & 0.664 & 0.936 \\
    \cline{3-9}
     &  & HJ & 0.710 & 0.970 & 0.722 & 0.972 & 0.672 & 0.946 \\
    \cline{2-9}
     & \multirow{2}{*}{$\alpha=2,\ \theta=2$} & KM & 0.550 & 0.968 & 0.650 & 0.967 & 0.819 & 0.953 \\
    \cline{3-9}
     &  & HJ & 0.556 & 0.970 & 0.651 & 0.973 & 0.823 & 0.955 \\
    \cline{2-9}
     & \multirow{2}{*}{$\alpha=2,\ \theta=5$} & KM & 0.164 & 0.944 & 0.186 & 0.949 & 0.266 & 0.968 \\
    \cline{3-9}
     &  & HJ & 0.168 & 0.940 & 0.186 & 0.948 & 0.263 & 0.959 \\
    \cline{2-9}
     & \multirow{2}{*}{$\alpha=5,\ \theta=1$} & KM & 0.543 & 0.977 & 0.546 & 0.970 & 0.442 & 0.938 \\
    \cline{3-9}
     &  & HJ & 0.530 & 0.974 & 0.519 & 0.967 & 0.430 & 0.942 \\
    \cline{2-9}
     & \multirow{2}{*}{$\alpha=5,\ \theta=2$} & KM & 0.319 & 0.947 & 0.370 & 0.980 & 0.534 & 0.967 \\
    \cline{3-9}
     &  & HJ & 0.315 & 0.958 & 0.366 & 0.977 & 0.525 & 0.962 \\
    \cline{2-9}
     & \multirow{2}{*}{$\alpha=5,\ \theta=5$} & KM & 0.016 & 0.948 & 0.017 & 0.930 & 0.023 & 0.916 \\
    \cline{3-9}
     &  & HJ & 0.017 & 0.954 & 0.016 & 0.934 & 0.024 & 0.922 \\
    \hline
    \multirow{12}{*}{50} & \multirow{2}{*}{$\alpha=2,\ \theta=1$} & KM & 0.697 & 0.972 & 0.743 & 0.974 & 0.718 & 0.952 \\
    \cline{3-9}
     &  & HJ & 0.680 & 0.968 & 0.743 & 0.970 & 0.706 & 0.945 \\
    \cline{2-9}
     & \multirow{2}{*}{$\alpha=2,\ \theta=2$} & KM & 0.504 & 0.964 & 0.610 & 0.971 & 0.798 & 0.959 \\
    \cline{3-9}
     &  & HJ & 0.501 & 0.972 & 0.606 & 0.971 & 0.803 & 0.962 \\
    \cline{2-9}
     & \multirow{2}{*}{$\alpha=2,\ \theta=5$} & KM & 0.134 & 0.929 & 0.152 & 0.958 & 0.218 & 0.980 \\
    \cline{3-9}
     &  & HJ & 0.134 & 0.932 & 0.152 & 0.947 & 0.220 & 0.973 \\
    \cline{2-9}
     & \multirow{2}{*}{$\alpha=5,\ \theta=1$} & KM & 0.495 & 0.966 & 0.537 & 0.972 & 0.476 & 0.951 \\
    \cline{3-9}
     &  & HJ & 0.491 & 0.962 & 0.545 & 0.969 & 0.489 & 0.967 \\
    \cline{2-9}
     & \multirow{2}{*}{$\alpha=5,\ \theta=2$} & KM & 0.273 & 0.959 & 0.323 & 0.975 & 0.482 & 0.976 \\
    \cline{3-9}
     &  & HJ & 0.272 & 0.954 & 0.323 & 0.975 & 0.479 & 0.972 \\
    \cline{2-9}
     & \multirow{2}{*}{$\alpha=5,\ \theta=5$} & KM & 0.011 & 0.961 & 0.010 & 0.929 & 0.014 & 0.930 \\
    \cline{3-9}
     &  & HJ & 0.011 & 0.945 & 0.011 & 0.934 & 0.014 & 0.930 \\
    \hline
  \end{tabular}%
  }

  \begin{minipage}{0.98\textwidth}
  \footnotesize
  \emph{Note.} The fixed censoring time \(C\) was determined by the target censoring proportion, as described in the simulation design. KM denotes the Kaplan--Meier plotting estimator and HJ denotes the Herd--Johnson plotting estimator.
  \end{minipage}
\end{table}

\bigskip

\begin{landscape}
\begin{table}[p]
\centering
\caption{Coverage probability and average interval length for $S(1.5)$ using the comparison methods under Type-I censoring (95\% confidence intervals; 2,000 Monte Carlo replications; 1,000 GPQ or bootstrap draws where applicable)}
\label{tab:type1_s15_transposed}
\fontsize{7.2}{8.1}\selectfont
\setlength{\tabcolsep}{1.8pt}
\renewcommand{\arraystretch}{0.94}
\textit{Entries are coverage probability (average interval length).}\par\vspace{3pt}
\begin{tabular}{@{}c*{12}{c}@{}}
\toprule
\multicolumn{13}{c}{\textbf{Panel: $n=20$}} \\
\addlinespace[1pt]
$(\alpha,\theta)$ & \multicolumn{4}{c}{20\% censoring} & \multicolumn{4}{c}{30\% censoring} & \multicolumn{4}{c}{50\% censoring} \\
\cmidrule(lr){2-5}\cmidrule(lr){6-9}\cmidrule(lr){10-13}
 & WLMA & B--LSE & B--MLE & A--MLE & WLMA & B--LSE & B--MLE & A--MLE & WLMA & B--LSE & B--MLE & A--MLE \\
\midrule
$(2,1)$ & 0.721 (0.151) & 0.960 (0.335) & 0.907 (0.248) & 0.854 (0.236) & 0.498 (0.117) & 0.958 (0.379) & 0.927 (0.283) & 0.844 (0.256) & 0.340 (0.085) & 0.958 (0.485) & 0.928 (0.385) & 0.798 (0.322) \\
$(2,2)$ & 0.719 (0.373) & 0.942 (0.348) & 0.933 (0.356) & 0.921 (0.351) & 0.449 (0.388) & 0.932 (0.365) & 0.915 (0.360) & 0.917 (0.354) & 0.049 (0.385) & 0.942 (0.448) & 0.936 (0.387) & 0.930 (0.387) \\
$(2,5)$ & 0.898 (0.244) & 0.918 (0.221) & 0.926 (0.169) & 0.885 (0.172) & 0.873 (0.279) & 0.936 (0.222) & 0.913 (0.170) & 0.864 (0.172) & 0.764 (0.374) & 0.934 (0.225) & 0.916 (0.174) & 0.872 (0.176) \\
$(5,1)$ & 0.819 (0.019) & 0.957 (0.151) & 0.893 (0.049) & 0.630 (0.017) & 0.740 (0.016) & 0.965 (0.205) & 0.896 (0.076) & 0.596 (0.028) & 0.586 (0.019) & 0.963 (0.328) & 0.927 (0.166) & 0.570 (0.065) \\
$(5,2)$ & 0.851 (0.346) & 0.924 (0.296) & 0.908 (0.272) & 0.890 (0.282) & 0.770 (0.382) & 0.926 (0.300) & 0.919 (0.276) & 0.906 (0.285) & 0.436 (0.467) & 0.937 (0.315) & 0.903 (0.281) & 0.899 (0.286) \\
$(5,5)$ & 0.910 (0.035) & 0.930 (0.050) & 0.918 (0.019) & 0.788 (0.012) & 0.905 (0.044) & 0.941 (0.054) & 0.933 (0.021) & 0.782 (0.013) & 0.891 (0.069) & 0.944 (0.058) & 0.903 (0.023) & 0.742 (0.014) \\
\addlinespace[2pt]
\midrule
\addlinespace[2pt]
\multicolumn{13}{c}{\textbf{Panel: $n=30$}} \\
\addlinespace[1pt]
$(\alpha,\theta)$ & \multicolumn{4}{c}{20\% censoring} & \multicolumn{4}{c}{30\% censoring} & \multicolumn{4}{c}{50\% censoring} \\
\cmidrule(lr){2-5}\cmidrule(lr){6-9}\cmidrule(lr){10-13}
 & WLMA & B--LSE & B--MLE & A--MLE & WLMA & B--LSE & B--MLE & A--MLE & WLMA & B--LSE & B--MLE & A--MLE \\
\midrule
$(2,1)$ & 0.545 (0.114) & 0.968 (0.287) & 0.905 (0.205) & 0.865 (0.204) & 0.318 (0.085) & 0.958 (0.336) & 0.925 (0.237) & 0.861 (0.224) & 0.141 (0.049) & 0.969 (0.451) & 0.928 (0.333) & 0.812 (0.290) \\
$(2,2)$ & 0.607 (0.310) & 0.950 (0.286) & 0.935 (0.291) & 0.934 (0.289) & 0.246 (0.324) & 0.944 (0.299) & 0.943 (0.294) & 0.941 (0.292) & 0.003 (0.322) & 0.944 (0.365) & 0.925 (0.318) & 0.922 (0.319) \\
$(2,5)$ & 0.919 (0.201) & 0.932 (0.181) & 0.936 (0.143) & 0.909 (0.149) & 0.850 (0.232) & 0.912 (0.181) & 0.937 (0.145) & 0.917 (0.151) & 0.674 (0.313) & 0.927 (0.181) & 0.919 (0.146) & 0.896 (0.152) \\
$(5,1)$ & 0.714 (0.009) & 0.954 (0.111) & 0.905 (0.033) & 0.676 (0.013) & 0.574 (0.007) & 0.949 (0.146) & 0.889 (0.048) & 0.662 (0.018) & 0.421 (0.006) & 0.951 (0.271) & 0.920 (0.110) & 0.601 (0.041) \\
$(5,2)$ & 0.830 (0.293) & 0.932 (0.245) & 0.925 (0.229) & 0.918 (0.237) & 0.698 (0.321) & 0.949 (0.243) & 0.919 (0.230) & 0.915 (0.237) & 0.276 (0.390) & 0.930 (0.251) & 0.927 (0.236) & 0.927 (0.239) \\
$(5,5)$ & 0.926 (0.023) & 0.931 (0.034) & 0.931 (0.014) & 0.812 (0.009) & 0.917 (0.027) & 0.924 (0.037) & 0.935 (0.015) & 0.817 (0.010) & 0.912 (0.044) & 0.943 (0.042) & 0.934 (0.017) & 0.794 (0.011) \\
\addlinespace[2pt]
\midrule
\addlinespace[2pt]
\multicolumn{13}{c}{\textbf{Panel: $n=50$}} \\
\addlinespace[1pt]
$(\alpha,\theta)$ & \multicolumn{4}{c}{20\% censoring} & \multicolumn{4}{c}{30\% censoring} & \multicolumn{4}{c}{50\% censoring} \\
\cmidrule(lr){2-5}\cmidrule(lr){6-9}\cmidrule(lr){10-13}
 & WLMA & B--LSE & B--MLE & A--MLE & WLMA & B--LSE & B--MLE & A--MLE & WLMA & B--LSE & B--MLE & A--MLE \\
\midrule
$(2,1)$ & 0.338 (0.087) & 0.945 (0.237) & 0.925 (0.164) & 0.893 (0.168) & 0.122 (0.058) & 0.951 (0.279) & 0.935 (0.190) & 0.891 (0.192) & 0.024 (0.026) & 0.948 (0.395) & 0.936 (0.268) & 0.869 (0.247) \\
$(2,2)$ & 0.400 (0.244) & 0.942 (0.222) & 0.940 (0.225) & 0.940 (0.225) & 0.065 (0.254) & 0.945 (0.232) & 0.940 (0.227) & 0.939 (0.227) & 0.000 (0.261) & 0.954 (0.288) & 0.940 (0.247) & 0.933 (0.248) \\
$(2,5)$ & 0.895 (0.155) & 0.923 (0.137) & 0.924 (0.111) & 0.898 (0.115) & 0.827 (0.181) & 0.925 (0.140) & 0.937 (0.116) & 0.914 (0.120) & 0.582 (0.249) & 0.924 (0.138) & 0.919 (0.118) & 0.905 (0.123) \\
$(5,1)$ & 0.591 (0.004) & 0.933 (0.072) & 0.915 (0.018) & 0.716 (0.007) & 0.412 (0.002) & 0.940 (0.099) & 0.916 (0.025) & 0.680 (0.009) & 0.208 (0.001) & 0.948 (0.195) & 0.938 (0.060) & 0.643 (0.021) \\
$(5,2)$ & 0.757 (0.234) & 0.941 (0.192) & 0.950 (0.182) & 0.947 (0.187) & 0.567 (0.255) & 0.923 (0.190) & 0.924 (0.183) & 0.929 (0.187) & 0.099 (0.311) & 0.940 (0.191) & 0.930 (0.186) & 0.928 (0.188) \\
$(5,5)$ & 0.933 (0.014) & 0.921 (0.022) & 0.938 (0.010) & 0.857 (0.007) & 0.921 (0.017) & 0.928 (0.024) & 0.945 (0.011) & 0.871 (0.008) & 0.919 (0.027) & 0.936 (0.028) & 0.938 (0.012) & 0.842 (0.009) \\
\bottomrule
\end{tabular}
\par\vspace{3pt}
\begin{minipage}{0.98\linewidth}
\fontsize{6.5}{7.2}\selectfont
\emph{Note.} B--LSE denotes Bootstrap--LSE, B--MLE denotes Bootstrap--MLE, and A--MLE denotes asymptotic MLE.
\end{minipage}
\end{table}
\end{landscape}

\subsection{Simulation results for the reliability function}

Tables~\ref{tab:type1_survival_t1_2}, \ref{tab:type1_s12_transposed},
\ref{tab:type1_survival_t1_5}, and \ref{tab:type1_s15_transposed} report the
coverage probabilities and average interval lengths for \(S(1.2)\) and
\(S(1.5)\). For the GLA procedures, interval lengths generally decreased with
\(n\) and increased with censoring. Coverage was close to 0.95 in most
configurations, with the largest departures occurring for \(n=20\) under 50\%
censoring. KM and HJ again produced very similar results; their differences
were usually only a few percentage points and diminished as \(n\) increased.

The boundary setting \((\alpha,\theta)=(5,5)\) is particularly informative:
\(S(1.2)=0.999204\) and \(S(1.5)=0.997573\). Despite these reliabilities being
very close to one, the GLA intervals remained well calibrated in most cases.
For \(n=20\) with 50\% censoring, KM and HJ coverage was 0.925 and 0.896 for
\(S(1.2)\), and 0.928 and 0.905 for \(S(1.5)\); by \(n=50\), the corresponding
coverages were 0.924 and 0.928 for \(S(1.2)\), and 0.930 for both methods for
\(S(1.5)\). Thus, KM and HJ behaved similarly even near the upper boundary,
although HJ was slightly less accurate in the smallest, most heavily censored
case.

The relative performance also varied at interior reliability values. At
\(n=50\), \((\alpha,\theta)=(5,2)\), and 20\% censoring, the intervals for
\(S(1.5)\) had coverage 0.950 and average length 0.182 for Bootstrap--MLE,
compared with 0.959 (0.273) for GLA--KM and 0.954 (0.272) for GLA--HJ.
Here, Bootstrap--MLE achieved nominal coverage with shorter intervals.

For the boundary setting \((\alpha,\theta)=(5,5)\), Bootstrap--LSE generally maintained coverage
between about 0.92 and 0.96. The asymptotic MLE intervals were the shortest,
with coverage ranging from 0.719 to 0.860 for \(S(1.2)\) and from 0.742 to 0.871
for \(S(1.5)\). Bootstrap--MLE could also improve on GLA: at \(n=50\) with
50\% censoring, its \(S(1.5)\) coverage was 0.938 with average length 0.012,
compared with coverage 0.930 and average length 0.014 for both GLA versions.
These results illustrate that relative coverage
and interval-length performance depends on the parameter configuration and
censoring level.

      \section{Examples}
  \label{sec:examples}

We illustrate the procedures with a complete ball-bearing dataset and a Type-I censored dataset. For each, we report Cramér--von Mises goodness-of-fit
results, 90\% and 95\% intervals from GLA-complete for complete data or
GLA--KM and GLA--HJ for censored data, together with WLMA,
Bootstrap--LSE, Bootstrap--MLE, and asymptotic MLE, and fitted reliability curves
with GLA pointwise confidence bands.

\indent\textbf{Example 1.} The first dataset was reported by \citet{Lawless2003} and has also been used in previous studies
of Weibull reliability inference \citep{Xiang2015}. The data consist of the failure
times of 23 ball bearings:
\[
\begin{split}
&17.88,\ 28.92,\ 33.00,\ 41.52,\ 42.12,\ 45.60,\ 48.48,\ 51.84,\ 51.96,\\
&54.12,\ 55.56,\ 67.80,\ 68.64,\ 68.64,\ 68.88,\ 84.12,\ 93.12,\ 98.64,\\
&105.12,\ 105.84,\ 127.92,\ 128.04,\ 173.40.
\end{split}
\]

The complete dataset consists of 23 observed failures, with no censoring. We therefore use GLA-complete, the
complete-data GPQ procedure with median-rank plotting positions; no
probability-matching equation is solved. To assess the null hypothesis that
the lifetimes follow a two-parameter Weibull distribution, we use the
complete-data Cramér--von Mises statistic. It measures the integrated squared
discrepancy between the empirical and fitted CDFs, providing an overall
assessment of distributional fit \citep{Stephens1974}. For ordered observations
\(x_{1:n}\le\cdots\le x_{n:n}\), it is
\[
W^2=\frac{1}{12n}+\sum_{i=1}^n
\left\{\widehat F(x_{i:n})-\frac{2i-1}{2n}\right\}^2,
\]
where \(\widehat F\) is the Weibull CDF fitted by LSE or MLE.
For each estimator, the \(p\)-value is calibrated using 1,000 parametric
bootstrap samples of size 23 from its fitted Weibull model, with the same
estimator refitted in every sample. Table~\ref{tab:gof_ballbearing} shows
similar LSE and MLE fits; both \(p\)-values exceed 0.05, providing insufficient evidence
against the Weibull model.

\begin{center}
\captionof{table}{Cramér--von Mises goodness-of-fit results for the ball-bearing dataset}
\label{tab:gof_ballbearing}
\begin{tabular}{lrrrr}
\toprule
Estimation method
& $\hat{\alpha}$
& $\hat{\theta}$
& Cramér--von Mises statistic
& $p$-value \\
\midrule
LSE & 2.248 & 80.972 & 0.0616 & 0.4136 \\
MLE & 2.102 & 81.879 & 0.0580 & 0.3926 \\
\bottomrule
\end{tabular}
\end{center}

\begin{center}
\captionof{table}{Confidence intervals for the Weibull parameters in the ball-bearing dataset}
\label{tab:ballbearing_parameter_intervals}

\small
\setlength{\tabcolsep}{5pt}
\renewcommand{\arraystretch}{1.08}

\begin{tabular}{llccc|ccc}
\toprule
& & \multicolumn{3}{c|}{90\% confidence interval}
& \multicolumn{3}{c}{95\% confidence interval} \\
\cmidrule(lr){3-5}\cmidrule(lr){6-8}
Method & Parameter
& Lower & Upper & Length
& Lower & Upper & Length \\
\midrule

\multirow{2}{*}{GLA-complete}
& $\alpha$
& 1.620 & 3.182 & 1.561
& 1.520 & 3.386 & 1.866 \\
& $\theta$
& 69.036 & 95.603 & 26.568
& 66.534 & 98.877 & 32.343 \\
\addlinespace

\multirow{2}{*}{WLMA}
& $\alpha$
& 1.630 & 3.197 & 1.567
& 1.526 & 3.414 & 1.888 \\
& $\theta$
& 68.754 & 95.611 & 26.857
& 66.592 & 98.311 & 31.719 \\
\addlinespace

\multirow{2}{*}{Bootstrap--LSE}
& $\alpha$
& 1.565 & 3.077 & 1.512
& 1.460 & 3.282 & 1.822 \\
& $\theta$
& 67.789 & 94.191 & 26.402
& 65.457 & 96.879 & 31.421 \\
\addlinespace

\multirow{2}{*}{Bootstrap--MLE}
& $\alpha$
& 1.713 & 2.943 & 1.231
& 1.618 & 3.139 & 1.521 \\
& $\theta$
& 68.111 & 96.189 & 28.078
& 65.605 & 98.436 & 32.831 \\
\addlinespace

\multirow{2}{*}{Asymptotic MLE}
& $\alpha$
& 1.561 & 2.643 & 1.081
& 1.458 & 2.746 & 1.288 \\
& $\theta$
& 67.732 & 96.025 & 28.293
& 65.022 & 98.735 & 33.713 \\

\bottomrule
\end{tabular}
\par\vspace{3pt}
\begin{minipage}{0.95\linewidth}
\footnotesize
\emph{Note.} GLA-complete denotes the proposed complete-data procedure using
median-rank plotting positions. Its 90\% and 95\% intervals use the same
10,000 GPQ draws. Lengths are computed before rounding the endpoints.
WLMA denotes the generalized pivotal procedure of Jia et al.
\end{minipage}
\end{center}

\begin{center}
\includegraphics[width=\textwidth]{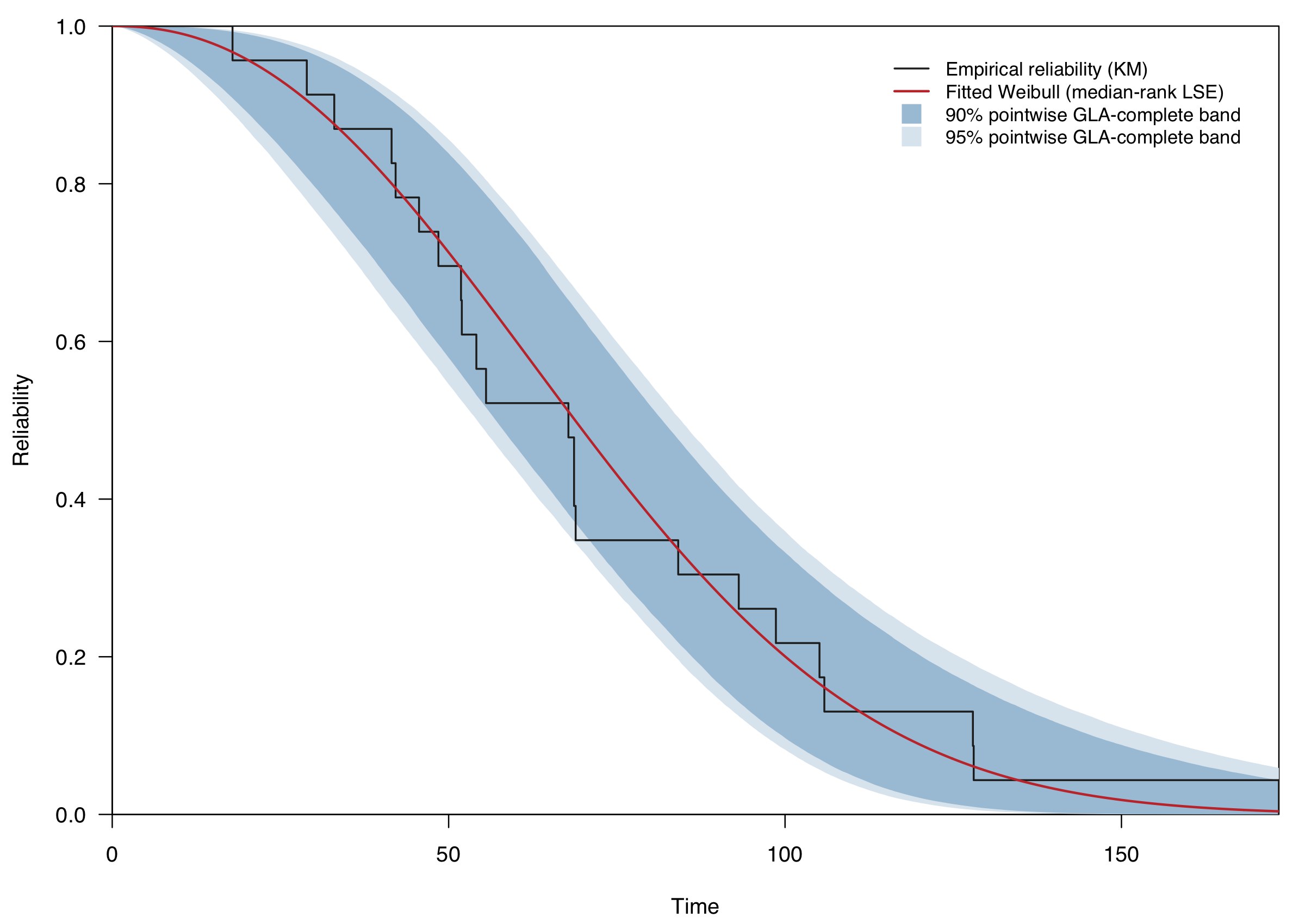}
\captionof{figure}{Empirical reliability estimate (KM), fitted Weibull curve using median-rank LSEs, and 90\% and 95\% pointwise GLA-complete confidence bands based on the same 10,000 GPQ draws used in Table~\ref{tab:ballbearing_parameter_intervals}. The darker and lighter regions denote the 90\% and 95\% bands, respectively.}
\label{fig:ballbearing_survival_gla}
\end{center}

Table~\ref{tab:ballbearing_parameter_intervals} shows close agreement among the
procedures. Bootstrap--MLE and
asymptotic MLE give shorter shape-parameter intervals, whereas scale-parameter
interval lengths are similar across methods. In
Figure~\ref{fig:ballbearing_survival_gla}, the fitted Weibull curve follows
the empirical reliability estimate closely, with pointwise confidence bands
displaying the uncertainty in reliability.

For reliability planning, the lower confidence band provides a conservative
assessment of component reliability at a selected operating time. The bands
describe uncertainty separately at each time; they do not provide simultaneous
coverage of the entire reliability curve.

\indent\textbf{Example 2.} The second dataset is taken from the NIST/SEMATECH e-Handbook of Statistical Methods \citep{nist_weibull_censored_example}
and consists of 20 lifetime observations:
\[
\begin{split}
&54,\ 187,\ 216,\ 240,\ 244,\ 335,\ 361,\ 373,\ 375,\ 386,\\
&500+,\ 500+,\ 500+,\ 500+,\ 500+,\ 500+,\ 500+,\ 500+,\ 500+,\ 500+.
\end{split}
\]

\indent Here, the notation \(500+\) indicates that the corresponding test unit survived beyond
time 500. Thus, ten failures were observed before the fixed censoring time \(C=500\), and
the remaining ten observations were Type-I right censored. The realized censoring proportion is
therefore 50\%. This example illustrates the principal setting considered in this article,
namely Weibull inference under conventional Type-I censoring.

We assess Weibull goodness of fit for the KM-based LSE, HJ-based LSE, and
MLE using the censored-data Cramér--von Mises statistic implemented in the R
package \texttt{GofCens} (version 1.5) \citep{GofCens2025,Besalu2025}.
The statistic compares the KM distribution estimate with the Weibull CDF
fitted by each estimator, accounting for Type-I right censoring. For each fit, the
\(p\)-value is calibrated using 1,000 samples from the fitted Weibull model,
censored at \(C=500\), with the same estimator refitted in every sample.
We use the package's bootstrap implementation for MLE and an
estimator-specific bootstrap implementation for the two LSE fits, retaining
the same statistic. All three results in Table~\ref{tab:gof_nist} provide
insufficient evidence against the Weibull model at the 5\% level.

\begin{center}
\begin{minipage}{\linewidth}
\centering
\captionof{table}{Cramér--von Mises goodness-of-fit results for the NIST dataset}
\label{tab:gof_nist}
\begin{tabular}{lrrrr}
\toprule
Estimation method
& $\hat{\alpha}$
& $\hat{\theta}$
& Cramér--von Mises statistic
& $p$-value \\
\midrule
LSE--KM & 1.490 & 598.673 & 0.432 & 0.331 \\
LSE--HJ & 1.478 & 627.228 & 0.431 & 0.396 \\
MLE & 1.721 & 606.550 & 0.429 & 0.406 \\
\bottomrule
\end{tabular}
\par\vspace{3pt}
{\footnotesize\raggedright
\emph{Note.} LSE--KM and LSE--HJ use Kaplan--Meier and Herd--Johnson plotting
estimates, respectively; MLE denotes maximum likelihood estimation.
Each \(p\)-value uses 1,000 bootstrap samples censored at \(C=500\), with the
corresponding estimator refitted in each sample. The fixed seeds are
20260723 for both LSE procedures and 20260724 for MLE.\par}
\end{minipage}
\end{center}


\begin{center}
\begin{minipage}{\linewidth}
\centering
\captionof{table}{Confidence intervals for the Weibull parameters in the NIST dataset}
\label{tab:nist_parameter_intervals}

\small
\setlength{\tabcolsep}{5pt}
\renewcommand{\arraystretch}{1.08}

\begin{tabular}{llccc|ccc}
\toprule
& & \multicolumn{3}{c|}{90\% confidence interval}
& \multicolumn{3}{c}{95\% confidence interval} \\
\cmidrule(lr){3-5}\cmidrule(lr){6-8}
Method & Parameter
& Lower & Upper & Length
& Lower & Upper & Length \\
\midrule

\multirow{2}{*}{GLA--KM}
& $\alpha$
& 1.133 & 3.518 & 2.386
& 0.992 & 3.912 & 2.920 \\
& $\theta$
& 263.883 & 1275.383 & 1011.499
& 246.513 & 1980.632 & 1734.119 \\
\addlinespace

\multirow{2}{*}{GLA--HJ}
& $\alpha$
& 1.102 & 3.459 & 2.357
& 0.976 & 3.785 & 2.809 \\
& $\theta$
& 264.495 & 1726.507 & 1462.012
& 249.975 & 3035.550 & 2785.575 \\
\addlinespace

\multirow{2}{*}{WLMA}
& $\alpha$
& 1.170 & 3.365 & 2.195
& 1.030 & 3.711 & 2.681 \\
& $\theta$
& 235.908 & 432.435 & 196.527
& 220.248 & 462.292 & 242.044 \\
\addlinespace

\multirow{2}{*}{Bootstrap--LSE}
& $\alpha$
& 0.705 & 2.423 & 1.718
& 0.641 & 2.779 & 2.139 \\
& $\theta$
& 421.286 & 1399.501 & 978.215
& 384.218 & 1795.665 & 1411.447 \\
\addlinespace

\multirow{2}{*}{Bootstrap--MLE}
& $\alpha$
& 1.116 & 3.031 & 1.915
& 0.970 & 3.268 & 2.299 \\
& $\theta$
& 449.972 & 942.865 & 492.893
& 431.638 & 1157.933 & 726.296 \\
\addlinespace

\multirow{2}{*}{Asymptotic MLE}
& $\alpha$
& 0.895 & 2.547 & 1.652
& 0.737 & 2.705 & 1.968 \\
& $\theta$
& 401.058 & 811.997 & 410.940
& 361.695 & 851.360 & 489.665 \\

\bottomrule
\end{tabular}
\end{minipage}
\par\vspace{3pt}
\begin{minipage}{0.95\linewidth}
\footnotesize
\emph{Note.} GLA--KM denotes the proposed generalized location--scale approach using the Kaplan--Meier plotting estimator, GLA--HJ denotes the proposed approach using the Herd--Johnson plotting estimator, and WLMA denotes the generalized pivotal procedure of Jia et al.
\end{minipage}
\end{center}

\begin{center}
\includegraphics[width=\textwidth]{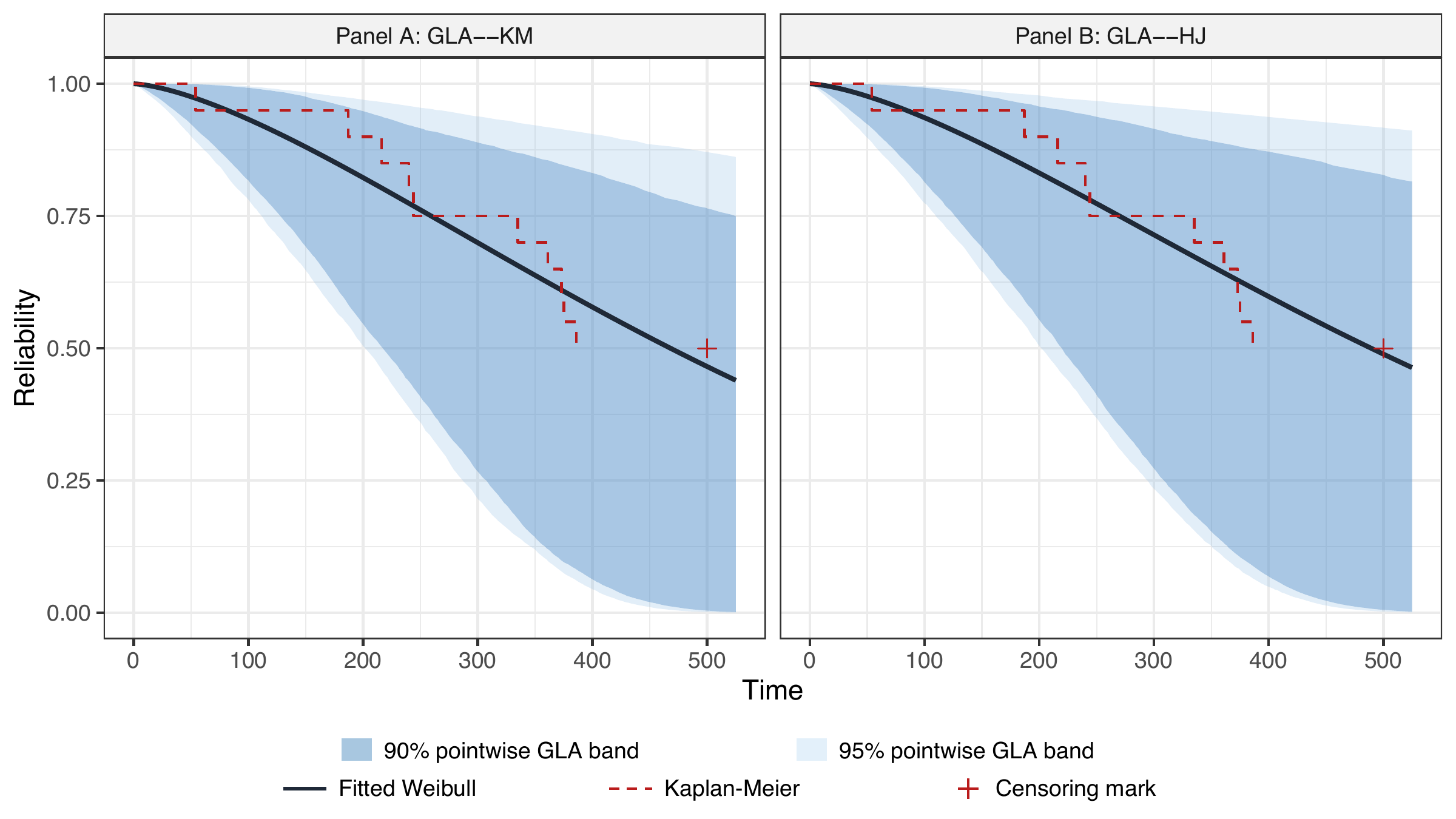}
\captionof{figure}{Kaplan--Meier estimate of the reliability function, fitted Weibull reliability curve, and 90\% and 95\% pointwise GLA confidence bands for the reliability function. Panels A and B are based on the Kaplan--Meier and Herd--Johnson plotting estimators, respectively. The plus symbols indicate censored observations at the fixed censoring time.}
\label{fig:nist_survival_gla}
\end{center}

Table~\ref{tab:gof_nist} reports \(p\)-values of 0.331, 0.396, and 0.406
for LSE--KM, LSE--HJ, and MLE, respectively; none of the three tests rejects
the Weibull model at the 5\% level. The intervals in
Table~\ref{tab:nist_parameter_intervals} are similar across methods for
\(\alpha\), but differ substantially for \(\theta\). GLA--KM, GLA--HJ, and
Bootstrap--LSE reflect considerable upper-tail uncertainty, whereas the much
shorter WLMA interval lies below all three point estimates. This difference
illustrates the coverage--length trade-offs observed in the Type-I
simulations.

Figure~\ref{fig:nist_survival_gla} shows that both GLA fits track the empirical
reliability estimate over the observed range. The pointwise confidence bands widen near and beyond the
censoring time because no failures occur after time 386 and half the sample is
censored at 500. This upper-tail uncertainty shows why interval width alone is
not sufficient under heavy censoring and, for maintenance or risk decisions
beyond \(C\), supports using a lower confidence bound or collecting additional
failure information rather than relying on the point estimate.

          \section{Discussion}
  \label{sec: discussion}
    \bigskip

In this article, we considered the problem of making generalized inference for the reliability
function of the two-parameter Weibull distribution under Type-I censoring. The
proposed methodology uses the Weibull-to-Gumbel transformation, under which the
Weibull model becomes a location--scale model. For complete samples, generalized
pivotal quantities follow directly from the equivariance of the LSEs. For
Type-I censored data, the observed failure count is used to obtain a
parameter-free Uniform order-statistic representation and to incorporate the
fixed censoring time into each GPQ draw. Appendix A establishes the
parameter-free and observed-value properties required for GPQs of the Weibull
shape, scale, and reliability parameters under the stated regularity
conditions.

The proposed GPQ methodology was compared with the WLMA generalized
pivotal procedure, parametric bootstrap intervals based on LSE and MLE, and
asymptotic MLE intervals through an extensive simulation study covering sample
sizes \(n\in\{20,30,50\}\), shape parameters
\(\alpha\in\{2,5\}\), scale parameters \(\theta\in\{1,2,5\}\), target
censoring proportions of 20\%, 30\%, and 50\%, and reliability at two prespecified times.
The selected shape values correspond to increasing-hazard reliability
settings. The simulation study indicates that the proposed method generally
achieves coverage probabilities close to the nominal level in many
scenarios considered. The
comparison methods show different coverage-length trade-offs across sample
sizes, censoring levels, and parameter settings; no method is uniformly best
under both criteria.

The practical applicability of the proposed methodology was illustrated using
real reliability data. The fitted reliability curves and pointwise confidence bands show how
the procedures quantify uncertainty relevant to component reliability and
maintenance decisions, particularly beyond the observed failure range.

A practical limitation is that each GPQ draw requires numerical
root finding for the auxiliary failure probability, which can be
computationally intensive. Nevertheless, the proposed procedure achieved
coverage close to the nominal level in many scenarios examined, including
several small-sample and heavily censored settings.

    \section*{Appendix A: Theoretical justification of the Gumbel-based GPQs}
\label{appendix}

\addcontentsline{toc}{section}{Appendix A. Theoretical justification of the Gumbel-based GPQs}
\begingroup
\renewcommand{\thesubsection}{1.\arabic{subsection}}
\setcounter{subsection}{0}
Throughout this appendix, the Weibull distribution is parameterized by the shape parameter
\(\alpha>0\) and scale parameter \(\theta>0\), with distribution function
\begin{equation}
F_X(x;\alpha,\theta)=1-\exp\left\{-\left(\frac{x}{\theta}\right)^\alpha\right\},
\qquad x>0.
\label{app:eq:weibull-cdf}
\end{equation}
We write
\begin{equation}
Y=\ln X,\qquad \nu=\ln\theta,\qquad \sigma=\frac{1}{\alpha}.
\label{app:eq:log-param}
\end{equation}
The corresponding standard Gumbel distribution used here is the standard minimum
extreme-value distribution with cumulative distribution function
\begin{equation}
F_0(z)=1-\exp\{-\exp(z)\},\qquad -\infty<z<\infty,
\label{app:eq:min-gumbel-cdf}
\end{equation}
and quantile function
\begin{equation}
q(u)=F_0^{-1}(u)=\ln\{-\ln(1-u)\},\qquad 0<u<1.
\label{app:eq:min-gumbel-quantile}
\end{equation}
This convention is fixed throughout the appendix. If the maximum Gumbel convention
\(\exp\{-\exp(-z)\}\) is used instead, the signs in the transformation and the regression
scores must be changed accordingly.

\subsection*{\textit{A.1 Weibull-to-Gumbel transformation}}
\label{app:subsec:weibull-gumbel}

As established in Section~\ref{Sec:2}, if \(X\sim W(\alpha,\theta)\), then
\(Y=\ln X=\nu+\sigma Z\), where \(\nu=\ln\theta\),
\(\sigma=1/\alpha\), and \(Z\sim F_0\). Consequently, for any prespecified
time \(t>0\),
\[
\alpha=\frac{1}{\sigma},\qquad \theta=\exp(\nu),\qquad
S_X(t)=\exp\left[-\exp\left\{\frac{\ln t-\nu}{\sigma}\right\}\right].
\]
These identities are used below to transform the GPQs for \(\nu\) and \(\sigma\)
back to the Weibull parameters and reliability function.

\subsection*{\textit{A.2 Complete-data GPQ justification}}
\label{app:subsec:complete-gpq}
Assume first that \(Y_1,\ldots,Y_n\) are completely observed and satisfy the Gumbel
location-scale model
\begin{equation}
Y_i=\nu+\sigma Z_i,
\qquad
Z_1,\ldots,Z_n
\stackrel{\mathrm{iid}}{\sim}
F_0,
\label{app:eq:complete-gumbel-model}
\end{equation}
where \(F_0\) is given in \eqref{app:eq:min-gumbel-cdf}. Since \(\sigma>0\), ordering is
preserved, so
\begin{equation}
Y_{i:n}
=
\nu+\sigma Z_{i:n},
\qquad
i=1,\ldots,n.
\label{app:eq:ordered-location-scale}
\end{equation}
Let \(p_1,\ldots,p_n\) be fixed plotting positions and define the corresponding Gumbel
scores
\begin{equation}
w_i =q(p_i) =
\ln\{-\ln(1-p_i)\},
\qquad
i=1,\ldots,n.
\label{app:eq:complete-scores}
\end{equation}

Write
\[
\bar w=\frac1n\sum_{i=1}^{n}w_i,
\qquad
D_n=\sum_{i=1}^{n}(w_i-\bar w)^2.
\]
The LSEs obtained from the regression of \(Y_{i:n}\) on \(w_i\) are
\begin{equation}
\widehat{\sigma}
= \frac{
\sum_{i=1}^{n} (w_i-\bar w) (Y_{i:n}-\bar Y)
}
{D_n},
\qquad
\widehat{\nu}
= \bar Y-\widehat{\sigma}\bar w,
\label{app:eq:complete-lse}
\end{equation}
where
\[
\bar Y=\frac1n\sum_{i=1}^{n}Y_{i:n}.
\]

For a standardized Gumbel sample
\[
\mathbf Z=(Z_1,\ldots,Z_n),
\]
define
\begin{equation}
\widehat{\sigma}(\mathbf Z)
=
\frac{
\sum_{i=1}^{n}
(w_i-\bar w)
(Z_{i:n}-\bar Z)
}
{D_n},
\label{app:eq:sigmaZ}
\end{equation}
where $$
\bar Z=\frac1n\sum_{i=1}^{n}Z_{i:n}.
$$

Also define
\begin{equation}
\widehat{\nu}(\mathbf Z)
=
\bar Z
-
\bar w\,\widehat{\sigma}(\mathbf Z).
\label{app:eq:nuZ}
\end{equation}

\noindent\textbf{Proposition 1.}
Under the complete-data model \eqref{app:eq:complete-gumbel-model},
\begin{equation}
\widehat{\sigma}
=
\sigma\widehat{\sigma}(\mathbf Z),
\label{app:eq:sigma-equiv}
\end{equation}
and
\begin{equation}
\widehat{\nu}
= \nu+\sigma\widehat{\nu}(\mathbf Z).
\label{app:eq:nu-equiv}
\end{equation}
Therefore, if
\(\widehat{\sigma}_{\mathrm{obs}}\) and
\(\widehat{\nu}_{\mathrm{obs}}\)
denote the observed LSEs, then
\begin{equation}
G_{\sigma}
=
\frac{
\widehat{\sigma}_{\mathrm{obs}}
}
{
\widehat{\sigma}(\mathbf Z)
},
\label{app:eq:Gsigma-complete}
\end{equation}
and
\begin{equation}
G_{\nu}
=
\widehat{\nu}_{\mathrm{obs}}
-
G_{\sigma}
\widehat{\nu}(\mathbf Z)
\label{app:eq:Gnu-complete}
\end{equation}
are generalized pivotal quantities for \(\sigma\) and \(\nu\), respectively.

\noindent\textit{Proof.}
Using \eqref{app:eq:ordered-location-scale},
\begin{align}
\widehat{\sigma}
&=
{
\sum_{i=1}^{n}
(w_i-\bar w)
\{\nu+\sigma Z_{i:n}
-(\nu+\sigma\bar Z)\}
} /
{D_n} \\
& =
\sigma
{
\sum_{i=1}^{n}
(w_i-\bar w)
(Z_{i:n}-\bar Z)
} /
{D_n} =
\sigma\widehat{\sigma}(\mathbf Z).
\end{align}

Similarly,
\begin{align}
\widehat{\nu}
&=
\bar Y-\widehat{\sigma}\bar w
=
\nu+\sigma\bar Z
-
\sigma\widehat{\sigma}(\mathbf Z)\bar w 
\\
& = \nu +
\sigma
\left\{
\bar Z-\bar w\,\widehat{\sigma}(\mathbf Z)
\right\} =
\nu+\sigma\widehat{\nu}(\mathbf Z).
\end{align}
Thus \eqref{app:eq:sigma-equiv} and
\eqref{app:eq:nu-equiv} hold.
Since \(\mathbf Z\) follows the standard minimum Gumbel distribution,
\(\widehat{\sigma}(\mathbf Z)\) and
\(\widehat{\nu}(\mathbf Z)\)
are free of unknown parameters. Furthermore, at the observed sample point,
\[
\widehat{\sigma}_{\mathrm{obs}}
=
\sigma\widehat{\sigma}(\mathbf z),
\qquad
\widehat{\nu}_{\mathrm{obs}}
=
\nu+\sigma\widehat{\nu}(\mathbf z),
\]
where
$
\mathbf z = (z_1,\ldots,z_n)
$
denotes the observed standardized sample. Hence
\[
G_{\sigma}
=
\frac{
\widehat{\sigma}_{\mathrm{obs}}
}
{
\widehat{\sigma}(\mathbf z)
}
=
\sigma,
\]
and
\[
G_{\nu}
=
\widehat{\nu}_{\mathrm{obs}}
-
\sigma\widehat{\nu}(\mathbf z)
=
\nu.
\]
Therefore \(G_{\sigma}\) and \(G_{\nu}\) satisfy the defining properties of generalized
pivotal quantities.
\hfill\(\square\)

By the substitution property of generalized pivotal quantities
\citep{Weerahandi2004}, the corresponding GPQs
for the original Weibull parameters are
\begin{equation}
G_{\alpha}
=
\frac{1}{G_{\sigma}},
\qquad
G_{\theta}
=
\exp\!\left\{
G_{\nu}
\right\}.
\label{app:eq:complete-weibull-gpqs}
\end{equation}
For a prespecified time \(t>0\), the GPQ for the reliability at time \(t\) is
\begin{equation}
G_{S(t)}
=
\exp\left[
-
\left(
\frac{
t
}{
G_{\theta}
}
\right)^{G_{\alpha}}
\right].
\label{app:eq:complete-survival-gpq}
\end{equation}

\subsection*{\textit{A.3 Type-I censored data: GPQ construction}}
\label{app:subsec:type1-censored-gpq}

Now suppose that the experiment is conducted under a Type-I censoring design with
predetermined censoring time \(C>0\). Let
\begin{equation}
c=\ln C,
\qquad
Y_i^{\mathrm{obs}}=\min(Y_i,c),
\qquad
\delta_i=I(Y_i\le c),
\qquad
R=\sum_{i=1}^{n}\delta_i.
\label{app:eq:censoring-setup}
\end{equation}

Once the Type-I censored sample is observed, its failure count \(R=r\) is
known. The construction below uses this observed count. This subsection considers
\(2\le r<n\); if \(r=n\), the complete-data construction in
Subsection~\ref{app:subsec:complete-gpq} is used. The case \(r<2\) does not provide
enough observed failures to define a slope-based LSE.

Define the standardized censoring point and its standard Gumbel probability by
\begin{equation}
z_c
=
\frac{c-\nu}{\sigma},
\qquad
p
=
F_0(z_c).
\label{app:eq:zc-p}
\end{equation}
Because \(c\) is fixed on the data scale, \(z_c\) and \(p\) depend on the unknown
parameters \(\nu\) and \(\sigma\).
The construction below fixes the observed number of failures \(R=r\) to obtain
the parameter-free reference representation needed for the GPQ derivation.
Let
\[
Y_{1:r}<\cdots<Y_{r:r}
\]
denote the ordered observed log failure times and let
\[
Z_{1:r}<\cdots<Z_{r:r}
\]
denote the corresponding ordered standardized failure times.

First, conditional on \(R=r\), define
\begin{equation}
U_{j:r}
=
\frac{F_0(Z_{j:r})}{p},
\qquad
j=1,\ldots,r.
\label{app:eq:U-transform}
\end{equation}
Then
\[
U_{1:r}<\cdots<U_{r:r}
\]
are the order statistics from \(r\) independent Uniform\((0,1)\) random variables.
Equivalently,
\begin{equation}
Z_{j:r}
=
q\{pU_{j:r}\},
\qquad
j=1,\ldots,r.
\label{app:eq:conditional-z-representation}
\end{equation}

\noindent\textbf{Proposition 2.}
Given the observed failure count \(R=r\), the representation
\eqref{app:eq:conditional-z-representation} holds with
\(U_{1:r},\ldots,U_{r:r}\) distributed as Uniform\((0,1)\) order statistics.
Their order-statistic distribution is free of \(\nu\) and \(\sigma\).

\bigskip
\noindent\textit{Proof.}
Given \(R=r\), the unordered standardized failure times are independent draws from the
standard Gumbel distribution truncated to \((-\infty,z_c]\). Hence each has conditional
density
\[
f_{Z\mid Z\le z_c}(z)
=
\frac{f_0(z)}{F_0(z_c)}
=
\frac{f_0(z)}{p},
\qquad
z\le z_c.
\]
Therefore, the joint density of the ordered failure values
\(Z_{1:r}<\cdots<Z_{r:r}\), conditional on \(R=r\), is

\[
r!\prod_{j=1}^{r}\frac{f_0(z_j)}{p},
\qquad
-\infty<z_1<\cdots<z_r\le z_c.
\]
Now set
\[
u_j=\frac{F_0(z_j)}{p}.
\]
Then
$ 0<u_1<\cdots<u_r<1, $
\[
z_j=q(pu_j),
\]
and
\[
\left|
\frac{dz_j}{du_j}
\right|
=
\frac{p}{f_0\{q(pu_j)\}}.
\]

Hence the joint density of
\((U_{1:r},\ldots,U_{r:r})\) is

\[
r!
\prod_{j=1}^{r}
\frac{f_0\{q(pu_j)\}}{p}
\prod_{j=1}^{r}
\frac{p}{f_0\{q(pu_j)\}}
=
r!,
\qquad
0<u_1<\cdots<u_r<1.
\]

This is precisely the joint density of Uniform\((0,1)\) order statistics and is
free of unknown parameters.
\hfill\(\square\)

\bigskip
Let $
w_1,\ldots,w_r $
be the fixed scores used in the censored-data LSE. 
These scores may be obtained from the Kaplan-Meier estimator, the
Herd-Johnson estimator, or any other pre-specified plotting-position rule. 
They are functions of \(n\), \(r\), and the ranks only.

Write
\[
\bar w_r
=
\frac1r
\sum_{j=1}^{r}w_j,
\qquad
D_r
=
\sum_{j=1}^{r}
(w_j-\bar w_r)^2.
\]
The censored-data LSEs are
\begin{equation}
\widehat{\sigma}_r
=
\frac{
\sum_{j=1}^{r}
(w_j-\bar w_r)
(Y_{j:r}-\bar Y_r)
}
{D_r},
\qquad
\widehat{\nu}_r
=
\bar Y_r-\widehat{\sigma}_r\bar w_r,
\label{app:eq:censored-lse}
\end{equation}
where
\[
\bar Y_r
=
\frac1r
\sum_{j=1}^{r}
Y_{j:r}.
\]
For
\[
\mathbf U
=
(U_{1:r},\ldots,U_{r:r}),
\]
define
\begin{equation}
\widehat{\sigma}(\mathbf U,p)
=
\frac{
\sum_{j=1}^{r}
(w_j-\bar w_r)
\{q(pU_{j:r})-\bar Q(\mathbf U,p)\}
}
{D_r},
\label{app:eq:sigmaUp}
\end{equation}
where
\[
\bar Q(\mathbf U,p)
=
\frac1r
\sum_{j=1}^{r}
q(pU_{j:r}),
\]
and
\begin{equation}
\widehat{\nu}(\mathbf U,p)
=
\bar Q(\mathbf U,p)
-
\bar w_r\widehat{\sigma}(\mathbf U,p).
\label{app:eq:nuUp}
\end{equation}

Then
\eqref{app:eq:conditional-z-representation} and
\eqref{app:eq:censored-lse}
imply
\begin{equation}
\widehat{\sigma}_r
=
\sigma\widehat{\sigma}(\mathbf U,p),
\qquad
\widehat{\nu}_r
=
\nu+\sigma\widehat{\nu}(\mathbf U,p).
\label{app:eq:censored-lse-decomp}
\end{equation}

Moreover, the censoring threshold satisfies
\begin{equation}
c
=
\nu+\sigma q(p).
\label{app:eq:c-threshold}
\end{equation}
Combining
\eqref{app:eq:censored-lse-decomp}
and
\eqref{app:eq:c-threshold}
gives
\begin{equation}
K_{\mathrm{obs}}
:=
\frac{
c-\widehat{\nu}_{r,\mathrm{obs}}
}
{
\widehat{\sigma}_{r,\mathrm{obs}}
}
=
\frac{
q(p)-\widehat{\nu}(\mathbf U,p)
}
{
\widehat{\sigma}(\mathbf U,p)
}.
\label{app:eq:Kobs-identity}
\end{equation}
For a generated vector
\[
\mathbf U^*
=
(U_{1:r}^*,\ldots,U_{r:r}^*),
\]
of Uniform\((0,1)\) order statistics, define
\(\widetilde p=\widetilde p(\mathbf U^*)\)
as the solution in \((0,1)\) of
\begin{equation}
K_{\mathrm{obs}}
=
\frac{
q(\widetilde p)
-
\widehat{\nu}(\mathbf U^*,\widetilde p)
}
{
\widehat{\sigma}(\mathbf U^*,\widetilde p)
}.
\label{app:eq:ptilde-equation}
\end{equation}

The underlying regularity conditions are 
\[
\widehat{\sigma}(\mathbf U^*,\widetilde p)\neq 0,
\]
and that the solution
\[
\widetilde p\in(0,1)
\]
exists and is unique. In implementation, the equation is solved numerically and uniqueness
is checked for each generated Monte Carlo sample.
\par\medskip
\noindent\textbf{Proposition 3.}
Under the Type-I censoring model, for fixed \(R=r\) and under the regularity
conditions stated above,
\begin{equation}
G_{\sigma}
=
\frac{
\widehat{\sigma}_{r,\mathrm{obs}}
}
{
\widehat{\sigma}(\mathbf U^*,\widetilde p)
},
\label{app:eq:Gsigma-censored}
\end{equation}
and
\begin{equation}
G_{\nu}
=
c-G_{\sigma}q(\widetilde p)
\label{app:eq:Gnu-censored}
\end{equation}
are generalized pivotal quantities for \(\sigma\) and \(\nu\), respectively, for
each admissible observed failure count \(r\).

\par\medskip
\noindent\textit{Proof.}

By Proposition 2, for an observed \(R=r\),
\[
\mathbf U^*
=
(U_{1:r}^*,\ldots,U_{r:r}^*)
\]
has the order-statistic distribution from \(r\) independent Uniform\((0,1)\) random
variables, which is free of \(\nu\) and \(\sigma\).

The quantity \(\widetilde p\) is obtained by solving
\eqref{app:eq:ptilde-equation}, whose inputs are \(\mathbf U^*\), the observed constant
\(K_{\mathrm{obs}}\), and the fixed scores \(w_1,\ldots,w_r\). Hence
\[
\widehat{\sigma}(\mathbf U^*,\widetilde p)
\]
and
\[
\widehat{\nu}(\mathbf U^*,\widetilde p)
\]
depend only on \(\mathbf U^*\) and observed constants. Therefore the conditional
distributions of \(G_{\sigma}\) and \(G_{\nu}\) are free of unknown parameters.

It remains to verify the observed-value property. At the observed sample point, let
\begin{equation}
u_{j,\mathrm{obs}}
=
\frac{
F_0\!\left\{
(y_{j:r,\mathrm{obs}}-\nu)/\sigma
\right\}
}
{p},
\qquad
j=1,\ldots,r,
\label{app:eq:uobs}
\end{equation}
where
\[
p
=
F_0\!\left(
\frac{c-\nu}{\sigma}
\right).
\]

By \eqref{app:eq:Kobs-identity}, the true value \(p\) satisfies
\eqref{app:eq:ptilde-equation} when
\[
\mathbf U^*
=
\mathbf u_{\mathrm{obs}}.
\]
Under uniqueness,
$ \widetilde p=p. $
Using \eqref{app:eq:censored-lse-decomp},
\begin{align}
G_{\sigma}
\Big|_{\mathbf U^*=\mathbf u_{\mathrm{obs}}}
&=
\frac{
\widehat{\sigma}_{r,\mathrm{obs}}
}
{
\widehat{\sigma}(\mathbf u_{\mathrm{obs}},p)
}
\\
&=
\frac{
\sigma\widehat{\sigma}(\mathbf u_{\mathrm{obs}},p)
}
{
\widehat{\sigma}(\mathbf u_{\mathrm{obs}},p)
}
= \sigma.
\end{align}
Also, using \eqref{app:eq:c-threshold},
\begin{align}
G_{\nu}
\Big|_{\mathbf U^*=\mathbf u_{\mathrm{obs}}}
=
c-\sigma q(p) =
\nu.
\end{align}

Thus, given the observed failure count \(R=r\), the reference distributions of
\(G_{\sigma}\) and \(G_{\nu}\) are free of unknown parameters. Together with the
observed-value property established above, this verifies the
generalized pivotal properties for \(\sigma\) and \(\nu\).
\hfill\(\square\)

The GPQs for the original Weibull parameters are obtained by substitution:
\begin{equation}
G_{\alpha}
=
\frac{1}{G_{\sigma}},
\qquad
G_{\theta}
=
\exp\{G_{\nu}\}.
\label{app:eq:censored-weibull-gpqs}
\end{equation}

For a prespecified time \(t>0\), the GPQ for the reliability at time \(t\) is
\begin{equation}
G_{S(t)}
=
\exp\left[
-
\left(
\frac{t}{G_{\theta}}
\right)^{G_{\alpha}}
\right].
\label{app:eq:censored-survival-gpq}
\end{equation}
The resulting generalized confidence interval is computed using the observed
failure count \(R=r\). Its Monte Carlo implementation repeatedly generates
\(\mathbf U^*\), solves \eqref{app:eq:ptilde-equation}, computes
\(G_{\sigma}\) and \(G_{\nu}\), and then applies
\eqref{app:eq:censored-weibull-gpqs} or
\eqref{app:eq:censored-survival-gpq} as required.

\endgroup

  \bibliographystyle{apalike}
  \bibliography{WeibullLSE.bib}
  \end{document}